\documentclass[apsprl,longbibliography,showpacs,floatfix,superscriptaddress,twocolumn]{revtex4-2}
\usepackage[usenames,dvipsnames]{color} 
\usepackage{graphicx}
\usepackage{amsfonts}
\usepackage[figuresright]{rotating}  
\usepackage{amssymb}
\usepackage{bm}
\usepackage{amsmath}
\usepackage{mathtools}
\usepackage{psfrag}
\usepackage{multirow}
\usepackage{tabularx}
\usepackage{textcomp}
\usepackage{units}
\usepackage{lipsum}
\usepackage{xr}
\usepackage{physics}
\usepackage{soul}
\usepackage{titlesec}
\usepackage{times}
\usepackage[colorlinks=true,citecolor=blue,linkcolor=magenta,hypertexnames=false]{hyperref}
\usepackage{enumitem}

\usepackage{subfigure}

\renewcommand{\selectlanguage}[1]{}

\DeclareMathAlphabet\mathbfcal{OMS}{cmsy}{b}{n}

\titlespacing*{\section}
{0pt}{1ex plus .25ex}{1ex plus 1ex}
\titlespacing*{\subsection}
{0pt}{1ex plus .25ex}{1ex plus 1ex}
\titlespacing*{\subsubsection}
{0pt}{1ex plus .25ex}{1ex plus 1ex}

\titleformat{\section}
 {\fontsize{10}{15}\bfseries }
  {\thesection}
  {1em}
  {}

\titleformat{\subsection}
  {\bfseries }
  {\thesubsection}
  {2em}
  {}

\titleformat{\subsubsection}
  {\itshape}
  {\thesubsubsection}
  {2em}
  {}

\makeatletter

\titleclass{\subsubsubsection}{straight}[\subsection]

\newcounter{subsubsubsection}[subsubsection]
\renewcommand\thesubsubsubsection{\thesubsubsection.\arabic{subsubsubsection}}

\titleformat{\subsubsubsection}
   {\normalfont\normalsize\itshape \centering}{\thesubsubsubsection}{1em}{}
\titlespacing*{\subsubsubsection}
{0pt}{1ex plus .3ex}{1ex plus .3ex}

\makeatletter
\renewcommand\paragraph{\@startsection{paragraph}{5}{\z@}%
  {3.25ex \@plus1ex \@minus.2ex}%
  {-1em}%
  {\normalfont\normalsize}}
  
\renewcommand\subparagraph{\@startsection{subparagraph}{6}{\parindent}%
  {3.25ex \@plus1ex \@minus .2ex}%
  {-1em}%
  {\normalfont\normalsize}}
\def\toclevel@subsubsubsection{4}
\def\toclevel@paragraph{5}
\def\toclevel@paragraph{6}
\def\l@subsubsubsection{\@dottedtocline{4}{7em}{4em}}
\def\l@paragraph{\@dottedtocline{5}{10em}{5em}}
\def\l@subparagraph{\@dottedtocline{6}{14em}{6em}}
\makeatother

\begin{document}

\title{The Weyl-Mott point:\ topological and non-Fermi liquid behavior\\from an isolated Green's function zero}

\author{R. Flores-Calderón}
\affiliation{Max Planck Institute for the Physics of Complex Systems, Nöthnitzer Strasse 38, 01187 Dresden, Germany}
\affiliation{Max Planck Institute for Chemical Physics of Solids, Nöthnitzer Strasse 40, 01187 Dresden, Germany}
\author{Chris Hooley}
\affiliation{Centre for Fluid and Complex Systems, Coventry University, Coventry CV1 2TT, United Kingdom}

\begin{abstract}
We present a model in which a Hatsugai-Kohmoto interaction is added to a system of fermions with a Weyl point in their non-interacting dispersion relation, and analyze its behavior as a function of the chemical potential.  We show that the  model exhibits a {\it Weyl-Mott point\/} --- a single isolated Green's function zero --- and that this implies an emergent non-Fermi-liquid state at the border of the metallic regime and a gapped topological state for the insulating one.  The Weyl-Mott point inherits the topological charge from the original Green's function pole, and is therefore naturally associated with a strongly correlated chiral anomaly.
 \end{abstract}
\maketitle


\textit{Introduction.}  The intersection of strong correlations and topology produces arguably the most exotic states of matter, with phenomena including symmetry protected topological phases, symmetry-enriched topological order, fractionalized symmetry operations, and fractionalized excitations. One can classify such phases as long-range entangled \cite{levin2009, swingle2011, maciejko2012, neupert2011, santos2011, levin2011exact, levin2012} or short-range entangled \cite{chen2013, gu2014, vishwanath20133d, wang2013bosonTI, burnell2014, kapustin2014symmetry, fidkowski2013, wang2014}. The first class has as prototypical examples non-interacting fermionic topological phases \cite{3DTI-exp2, kane2005z2, kane2005quantum, Fu-Kane-Mele, ten-fold-way, Moorehomotopy2007}, which can further be classified as gapped or gapless. Much work has been done on extending these phases to the case where interactions are present; these are now known as symmetry protected topological phases (SPTs) \cite{haldane1983, affleck1988, chen2011_gap1Dspin, senthil2015} and gapless SPTs \cite{Frank-Ruben-1d, Ruben-TopoCritQuant, scaffidi2017,Thorngren-PRB-2021}. Many open questions remain regarding their connections with long-range entangled states via gauging, gauge-Higgs theories \cite{verresen2024higgs,chung2024higgs}, and many-body topological invariants in terms of single-particle Green's functions. There has been a recent surge in understanding the role of Green's function zeros in topological responses and fractionalization \cite{sachdev_chern_2023, mai2023,bollmann_topological_2023,wagner_mott_2023,setty_electronic_2023,setty_symmetry_2023,blason_unified_2023}.

As noted in ref.\ \cite{sachdev_chern_2023}, for 2D systems the many-body Chern number matches the Ishikawa-Matsuyama Green's function invariant when Luttinger's theorem is satisfied and there are no self-energy poles at the Fermi level. Fractional quantum Hall effects (FQHE) are then encoded in quantities explicitly dependent on Green's function zeros.  Meanwhile, ref.\ \cite{bollmann_topological_2023} showed in a $\mathbb{Z}_2$ gauged model that Green's function zeros determine the Ishikawa-Matsuyama topological invariant.  Ref.\ \cite{wagner_mott_2023} similarly found that the topology of the zeros is related to emergent spinons, which appear as edge states.

Some authors have argued for the relevance of exactly solvable models based on the Hatsugai-Kohmoto interaction \cite{hatsugai1992}. This has the form of the Hubbard interaction, except that it is diagonal in the momentum rather than the position basis. Such an interaction might seem unphysical due to its non-local properties; however, it has been argued \cite{phillips_exact_2020,zhao_proof_2023,huang_discrete_2022} that it represents a fixed-point for strongly-correlated electronic phases, making it a tractable model with the same long-wavelength physics as the usual Hubbard term. Ref.\ \cite{mai2023} uses this to study a topological Mott insulator, obtaining similar results to those of the abovementioned local theories.

 \begin{figure}[ht!]
    \centering\includegraphics[width=\columnwidth]{
    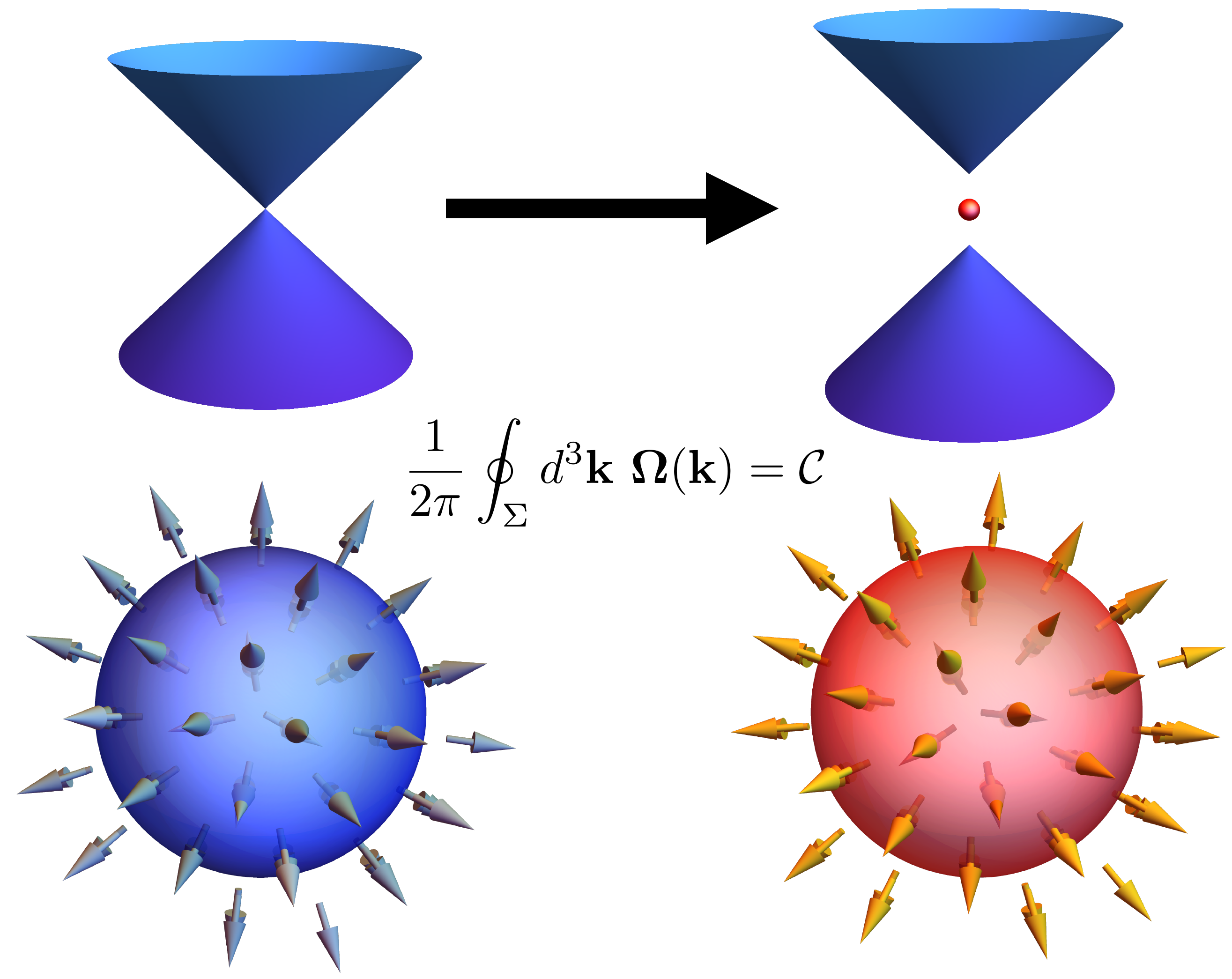}
    \caption{Schematic diagram of the transition from a Weyl semimetal (left) to a {\it Weyl-Mott point}, i.e.\ a single isolated zero of the Green's function (right). Upper panel:\ the band structure, showing the non-interacting limit (left) with a single pole at zero frequency, and the interacting case (right) which exhibits upper and lower Hubbard bands in addition to a single Green's function zero (shown in red).  Lower panel:\ the topological charge of the Weyl point remains invariant, but goes from being associated with a single Weyl point, i.e.\ a Green's function pole (left), to being associated with a Weyl-Mott point, i.e.\ an isolated Green's function zero (right).}
    \label{fig:schem}
 \end{figure}

In this paper we focus on the simplest gapless topological phase in 3D, a Weyl semimetal.  In the non-interacting limit a Weyl semimetal is characterized by a topological charge associated to the integral of Berry curvature around a diabolical point in momentum space, called a Weyl point. Physically observable effects appear in the coupling to electromagnetic fields, since the system exhibits a chiral anomaly.  Furthermore, when a boundary is introduced, Fermi-arc states appear connecting the Weyl points \cite{armitage_weyl_2018}. 

The strongly-correlated Weyl semimetal has been studied previously via numerical and approximate analytical methods. S.-C.\ Zhang and collaborators \cite{wang_chiral_2013} used mean-field methods to argue that under Hubbard interactions a gap is opened and a charge density wave (CDW) forms; in this scenario, not only does the chiral anomaly survive, but it is upgraded to a fully dynamical axion term via the CDW's fluctuating phase. Similar results have been predicted with other approaches 
\cite{zyuzin_topological_2012,burkovCDW_2020, zyuzin_topological_2012, crippa_nonlocal_2020, bobrow_monopole_2020}. Ref.\ \cite{burkovCDW_2020} indicates a connection between such CDWs and the 3D fractional quantum Hall effect. $(\text{TaSe}_4 )_2\text{I}$ has shown signs of a strongly-correlated CDW arising from a Weyl semimetal \cite{shi_charge-density-wave_2021}, and recent experiments have focused on gapless strongly-correlated Weyl-Kondo semimetals without a quasiparticle description \cite{kirschbaum_how_2023,hu_topological_2022}.

In this paper, we study the phase diagram of an exactly solvable lattice model of a strongly-correlated Weyl semimetal. Our Hatsugai-Kohmoto-Weyl (HKW) model is a lattice completion of one  first introduced by Morimoto and Nagaosa \cite{morimoto_weyl_2016}, who found that at half-filling there is a spectral Mott gap, making it a topological Weyl-Mott insulator, but one that still displays a nonvanishing Hall conductance and Fermi arcs. Here we study the model for all values of the chemical potential, and show how the defining non-zero topological charge is intimately tied to the presence of the isolated Green's function zero, found in ref.~\cite{morimoto_weyl_2016}, which we term a {\it Weyl-Mott point}. Indeed, the topological charge of the non-interacting Weyl point gets transferred from that model's Green's function pole to a Green's function zero (see Fig.~\ref{fig:schem}). The chiral anomaly previously predicted via a mean-field treatment \cite{zyuzin_topological_2012} can thus be reinterpreted even in the strongly-interacting limit as arising from the Weyl-Mott point. 
 
\textit{Model.} The Hamiltonian of our HKW model is
\begin{equation}
    H= \sum_{\textbf{k},\sigma,\sigma'} h_{{\bf k}\sigma\sigma'} c_{\textbf{k}\sigma}^\dagger c_{\textbf{k}\sigma'}^{\phantom{\dagger}}
    +U \sum_{\textbf{k}}n_{\textbf{k}\uparrow}n_{\textbf{k}\downarrow}; \label{hWHK}
\end{equation}
$c_{\textbf{k}\sigma}^\dagger$ creates an electron with momentum $\textbf{k}$ and spin projection $\sigma$, and  $n_{\textbf{k}\sigma} \equiv c_{\textbf{k}\sigma}^\dagger c^{\phantom{\dagger}}_{\textbf{k}\sigma}$ counts the number of electrons in the state $\textbf{k}\sigma$. The non-interacting dispersion is $h_{{\bf k}\sigma\sigma'}=\vec{d}_{\textbf{k}}\cdot \vec{\sigma}-\mu$, with $\vec{d}_{\textbf{k}}$ a three-vector, $\vec{\sigma}$ a vector of Pauli matrices in spin space, and $\mu$ a chemical potential controlling the filling.

It is useful to change to a basis where the non-interacting term is diagonal in spin space. We define $\alpha_{\textbf{k}s}=\sum_{\sigma}V_{s\sigma}(\textbf{k}) c_{\textbf{k}\sigma}$ as the new fermionic annihilation operators; here $s=\pm$, and the unitary matrix $V(\textbf{k})$ is chosen so that the first term becomes $\sum_{\textbf{k},s} \lambda_{{\bf k}s}\alpha^\dagger_{\textbf{k}s}\alpha^{\phantom{\dagger}}_{\textbf{k}s}$, with $\lambda_{\textbf{k}s}=-\mu+sd_{\textbf{k}}$ and $d_{\textbf{k}}=\vert {\vec{d}}_{\textbf{k}} \vert$. 
The Hatsugai-Kohmoto term retains its form under this basis change (see Supplemental Material).
For the $\alpha$ fermions we write it as $U\sum_{\textbf{k}}m_{\textbf{k}+}m_{\textbf{k}-}$, with $m_{\textbf{k}s}=\alpha^\dagger_{\textbf{k}s}\alpha^{\phantom{\dagger}}_{\textbf{k}s}$.

\begin{figure}[ht!]
\centering
\includegraphics[scale=0.4]{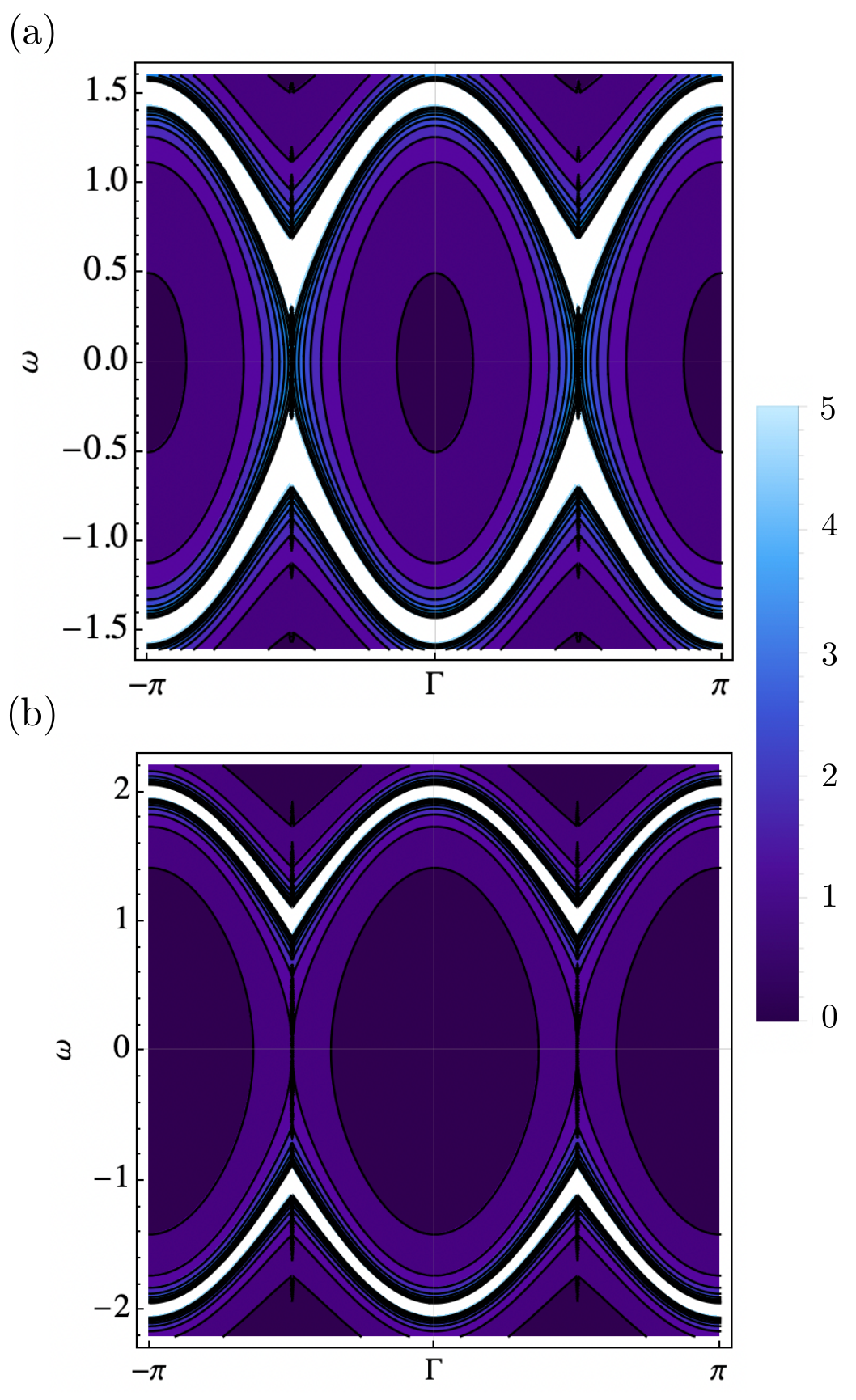}
\caption{The magnitude of the Green's function $\vert \kern-1pt \det G_{\sigma\sigma'}(\textbf{k},\omega) \vert$ as a contour plot in a frequency-momentum plane.  The momentum cut is taken along $\textbf{k}=(0,0,k_z)$; the Weyl-Mott points are at $\pm\textbf{k}_0=(0,0,\pm\pi/2)$. A divergence (white) indicates a pole of the single-particle Green's function; in addition to these, there is an exact isolated zero of the Green's function at $\omega=0$ for half-filling of \eqref{hWHK}.  Parameter values:\ $\beta=1000$, $M=2$, and (a) $U=1,\mu=1/2$; (b) $U=2,\mu=1$.}
\label{fig:detg}
\end{figure}

\textit{Green's function and Weyl-Mott points.} We now specialize to the case where the non-interacting dispersion relation has a Weyl point.  We take $\vec{d}_{\textbf{k}}=t\big(\sin k_x,\sin k_y,\cos k_x +\cos k_y +\cos k_z -M \big)$; for $1<M<3$ it describes a Weyl semimetal with Weyl nodes at $\pm \textbf{k}_0=(0,0,Q_z)$, where $\cos Q_z = 2-M$ \cite{armitage_weyl_2018} and the hopping matrix element $t>0$.

We consider first the Green's function at finite temperature $T=1/\beta$, in the Matsubara formalism with the imaginary time $\tau$. The Matsubara Green's function in the $s$-basis,  $\tilde{G}_{\textbf{k}s}(\tau)=-\langle \hat {\mathcal{T}} \alpha^{\phantom{\dagger}}_{\textbf{k}s}(\tau)\alpha^\dagger_{\textbf{k}s}(0) \rangle$, may be straightforwardly evaluated (see Supplemental Material); the result is
\begin{align}
     \tilde{G}_{s}(\textbf{k},i\omega_n)=\dfrac{1-\expval{m_{\textbf{k}\bar{s}}}}{i\omega_n-\lambda_{\textbf{k}s}}+\dfrac{\expval{m_{\textbf{k}\bar{s}}}}{i\omega_n-(\lambda_{\textbf{k}s}+U)},
\end{align}
where $\langle m_{{\bf k}s} \rangle$ is the average occupation of $\alpha$-mode ${\bf k}s$, and ${\bar s}$ denotes the opposite spin mode to $s$. The Matsubara Green's function in the original spin basis is thus given by 
$G_{\sigma\sigma'}(\textbf{k},i\omega_n) = \sum_{s} \left[ V^\dagger({\bf k}) \right]_{\sigma s}\tilde{G}_s(\textbf{k},i\omega_n)V_{s \sigma'}(\textbf{k})$.

The structure of this Green's function is very similar to that of ref.\ \cite{phillips_exact_2020}, and to the atomic-limit case:\ a two-pole structure reminiscent of the upper and lower Hubbard bands. The main difference is the spin index in the dispersion $\lambda_{{\bf k}s}$, which makes the Green's function behave differently for the two spin modes. The spectral function develops a gap between the $s=\pm$ bands, as shown schematically in the upper panel of Fig.~\ref{fig:schem}. Such a symmetric mass generation is consistent with previous results based on numerics or analytical approximations \cite{wang_chiral_2013, zyuzin_topological_2012, crippa_nonlocal_2020, bobrow_monopole_2020}, and with experiments \cite{shi_charge-density-wave_2021}.

We can see the spectral behavior by plotting $\vert \kern-1pt \det G(\textbf{k},\omega) \vert$ as in Fig.~\ref{fig:detg}, where we have specialized to half-filling. The upper and lower figures are for different strengths of the interaction $U$, but show qualitatively the same behavior, including a gap in frequency between the upper and lower bands.  The poles (white) are at $\lambda_{\textbf{k}s}$ and $\lambda_{\textbf{k}s}+U$, and there is a line connecting the upper and lower branches along which the Green's function is real-valued. It is easy to see from the functional form of the occupation number that if $\mu=U/2$ then $\expval{n_{\textbf{k}}} \equiv \sum_{\sigma} \expval{n_{\textbf{k}\sigma}}=1$. The spin-split occupation of the $\alpha$ modes is then given by
\begin{align}
\expval{m_{\textbf{k}s}}=\dfrac{1+e^{-\beta \lambda_{\textbf{k}s}}}{2+e^{-\beta\lambda_{\textbf{k}-}}+e^{-\beta\lambda_{\textbf{k}+}}}.
\label{avocc}
\end{align}
If $\textbf{k}\neq \pm \textbf{k}_0$ it is possible to take the zero temperature limit (see Supplemental Material) and conclude that $\expval{m_{\textbf{k}+}}=0$ while $\expval{m_{\textbf{k}-}}=1$. Because of this, the two-pole structure at half-filling gets reduced to a single pole for each spin index. Nevertheless, as we show next, the ground state of the system is still not adiabatically connected to the non-interacting limit.

This is due to the presence of Weyl nodes in the original non-interacting dispersion. At these points $d_{\textbf{k}}=0$; plugging this into the half-filled occupation leads to $\expval{m_{\textbf{k}+}}=\expval{m_{\textbf{k}-}}=1/2$. Surprisingly, then, for $\textbf{k}=\pm\textbf{k}_0$ we recover the two-pole structure seen in the Mott-insulating case. Moreover, these occupation numbers are the same as long as $0<\mu<U$, as shown in the Supplemental Material; the electronic system is thus not smoothly connected to the non-interacting limit. We can understand this from the self-energy $\Sigma^R(\textbf{k}_0,\omega)$ which  diverges for any value of $U$ as $\omega\rightarrow 0$, as can be seen from the real-time Green's function form
\begin{align}
\tilde{G}_s(\textbf{k}_0,\omega)= \dfrac{1}{\omega+i0^+-\dfrac{(U/2)^2}{\omega+i0^+}}. \label{gf}
\end{align}
Furthermore, the chemical potential at half-filling and $\mu=U/2$ always touches the upper half of one of the split Weyl points, as can be seen from the spectral function $\tilde{A}_s(\textbf{k},\omega) \equiv -2\Im G^R(\textbf{k},\omega)$ at the Weyl point:
\begin{align}
\tilde{A}_s(\textbf{k}_0,\omega) =\pi \left( \delta(\omega+U/2)+\delta(\omega-U/2)\right).
\end{align}

We see from (\ref{gf}) that $\tilde{G}_s(\textbf{k}_0,\omega)\rightarrow 0 $ as $\omega\rightarrow 0$, i.e.\ that the two-pole Mott structure leads to a \textit{single isolated Green's function zero}.  This was noted in ref.~\cite{morimoto_weyl_2016}, though it has broader consequences than those discussed there --- see below. We define a \textit{Weyl-Mott} point as a point in momentum space where an isolated Green's function zero appears in the gap formed by the splitting of a Weyl point in the Green's function poles, as in Fig.~\ref{fig:detg}.

For $\mu=U/2$ and $T\neq0$ the occupation number $\expval{m_{\textbf{k}+}}$ is non-zero for any $U$ (see Supplemental Material). The finite-temperature Green's function thus displays the two-pole structure even for $\textbf{k}\neq \textbf{k}_0$. The Green's function zero is precisely at $\omega_0=0$ for $\mu=U/2$; for generic $0<\mu<U$ it appears at $\omega_0=\frac{U}{2}-\mu$. We see from (\ref{avocc}) that for $0<\mu< U$ we always have half filling of the Weyl nodes at zero temperature; this proves that the Weyl-Mott point is stable within a range of the chemical potential and interaction strength. 

\textit{Topological invariant and phase diagram.} What is the topological nature of the Weyl-Mott points?  We will show that the topological charge of the Weyl node is preserved, even though the state is not adiabatically connected to the non-interacting limit.

Our starting point is again the single-particle Green's function, which we now analyze at zero frequency, taking the inverse to define an effective Hamiltonian in the spirit of refs.~\cite{skolimowski_luttingers_2022,blason_unified_2023}:\ $H^*(\textbf{k})=-G^{-1}_{\sigma\sigma'}(\textbf{k},i\omega_n\rightarrow 0 ) = -\sum_{s}V_{\sigma s}(\textbf{k})^\dagger\tilde{G}^{-1}_s(\textbf{k},i\omega_n\rightarrow 0 )V_{s \sigma'}(\textbf{k})$. For $T=0$ and half-filling the Green's function has a one-pole structure for each $s=\pm$ sector as long as we stay away from the Weyl point. Thus if we define a Gaussian surface $\Sigma$ encircling the Weyl point $\textbf{k}_0$ the effective Hamiltonian has the form
\begin{align}
    H^*(\textbf{k})= V(\textbf{k})^\dagger \left[
    \begin{pmatrix}
        d_{\textbf{k}} && 0 \\
        0 && -d_{\textbf{k}} 
        \end{pmatrix}+
        \begin{pmatrix}
       \mu && 0 \\
        0 && \mu-U 
        \end{pmatrix}\right] V(\textbf{k}).
\end{align}
We can now calculate the topological charge enclosed by $\Sigma$ by integrating the Berry curvature $\bm{\Omega}(\textbf{k})$ calculated from $H^*(\textbf{k})$. For a non-interacting Weyl point this gives
\begin{align}
    \dfrac{1}{2\pi}\oint_\Sigma \dd^3 \textbf{k}\  \bm{\Omega}(\textbf{k})= C,
\end{align}
since we have a well defined gap on the surface $\Sigma$. In the interacting case, $U>0$, the gap in $H^*(\Sigma)$ will increase, but the momentum dependence of the energies and the eigenfunctions will be the same as for the non-interacting Weyl point. We thus conclude that the topological charge of the Weyl-Mott points will still be quantized, and in our case given by $C=\pm1$, a result consistent with the Hall conductivity calculation from ref.~\cite{morimoto_weyl_2016}.

We thus see the spectral gap and a divergent self-energy accompanied by a topological charge at the Weyl-Mott point for $0\leq \mu \leq U$, not just at $\mu=U/2$ as considered in ref.~\cite{morimoto_weyl_2016}.  Since the Weyl cone was split in frequency, these topological properties must have a contribution from the Green's function zero. The Green's function topological invariant will indeed depend upon such points, but may fail to be the same as the non-zero many-body Chern number, as shown in ref.~\cite{yang_manifestation_2019}, and produce no surface states \cite{meng_unpaired_2019}. In our case, away from the Weyl point we have a non-interacting response, and we thus find that the many-body Chern number and the topological invariant are the same, as found in ref.~\cite{morimoto_weyl_2016}.

The same argument away from the Weyl point shows that any model with linearly dispersing Green's function zeros, which might be expected from previous studies \cite{wagner_mott_2023}, is not adiabatically connected to our model.  The single isolated Green's function zero instead heralds the appearance of a non-Fermi liquid state with a non-trivial topological invariant.

\begin{figure}[ht!]
\centering\includegraphics[width=\columnwidth]{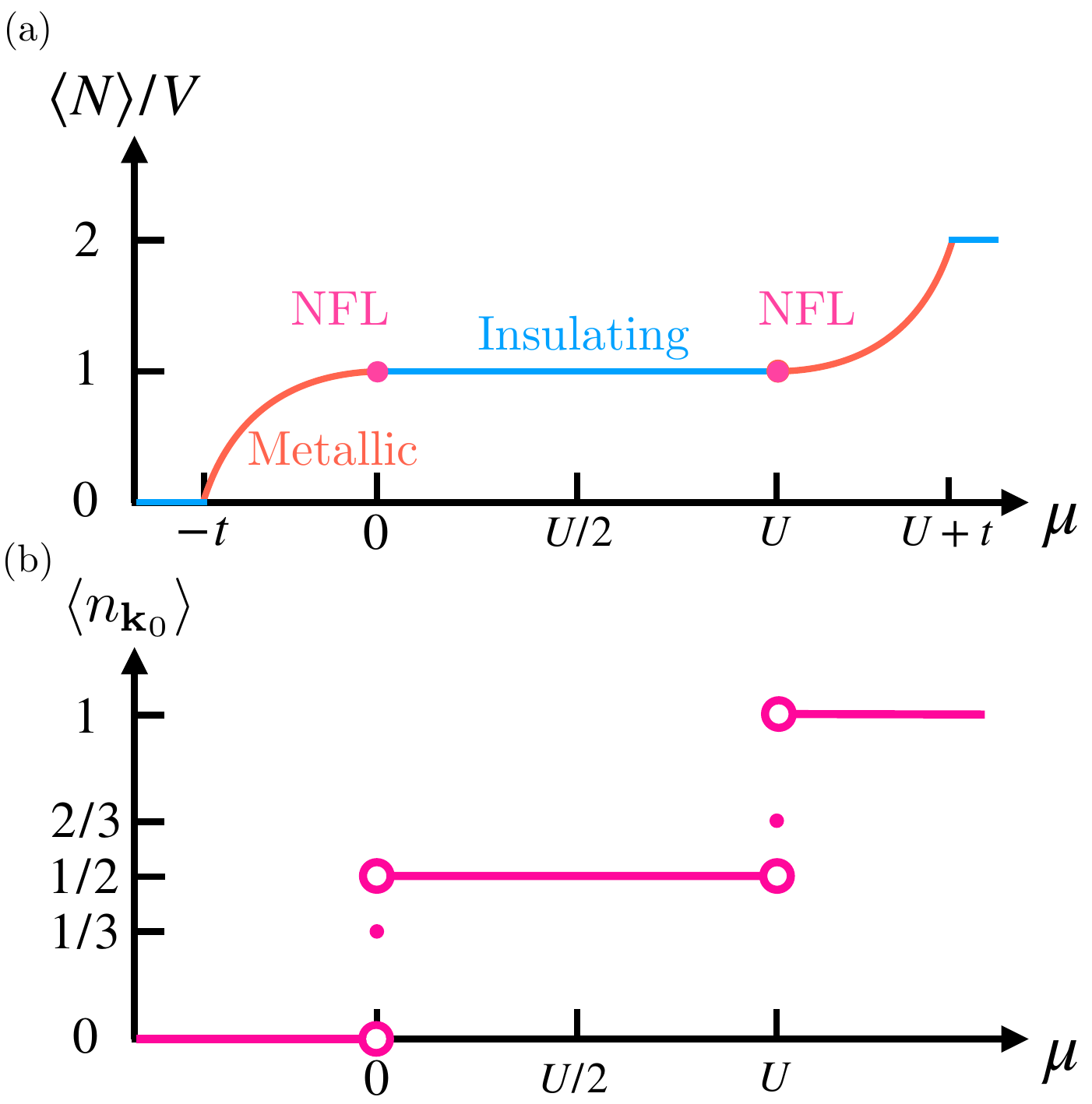}
    \caption{(a) Schematic phase diagram showing the average occupation per site at zero temperature as a function of chemical potential. Red and pink regions are metallic (i.e.\ have spectral weight at $\omega=0$); blue regions are insulating. The chemical potential crosses the Weyl-Mott point at $\mu=U/4$. (b) The average occupation in the lower band at one Weyl-Mott momentum $\textbf{k}=\textbf{k}_0$ as a function of chemical potential. Note the extended region where the value of the occupation is locked to $1/2$.}
    \label{fig:phase}
\end{figure}
Fig.~\ref{fig:phase} shows the phase diagram of our model and the Weyl-point occupation as a function of the chemical potential.  As long as $0<\mu<U$ the occupation stays at half-filling, with one electron per site. This agrees with the Luttinger count calculated using $\sum_{\textbf{k}s}\Theta[\Re(\tilde{G}_s(\textbf{k},0))]$. For $\mu\geq U$ the occupation goes smoothly to two electrons per site as all bands get filled --- see the Supplemental Material for details. We call the region  $U<\mu<U+t$ metallic as it behaves like a usual Fermi liquid with varying occupation and spectral weight at zero frequency.

Our model has a particle-hole symmetry:\ $\alpha^{\phantom{\dagger}}_{\textbf{k}s}\rightarrow \alpha^\dagger_{\textbf{k}\bar{s}}$ plus $\mu\rightarrow U-\mu$. Hence in the hole-doped case ($\mu<0$) the original Weyl points start to nucleate hole pockets where the occupation goes to zero. We again have a continuous interpolation, now to zero occupation.  For $\mu\leq -W$, where $W$ is the maximum value of $d_{\textbf{k}}$, the system has zero occupation.

{\it Weyl-Mott non-Fermi liquid.} The boundary points between the metallic and insulating phases, depicted in pink in Fig.~\ref{fig:phase}, we identify as non-Fermi liquid (NFL). Two facts point to such a classification. First, there is spectral weight at zero frequency; second, the Green's function has a two-pole structure which is not adiabatically connected to the non-interacting limit.

\begin{figure}[ht!]
\centering\includegraphics[width=0.9\columnwidth]{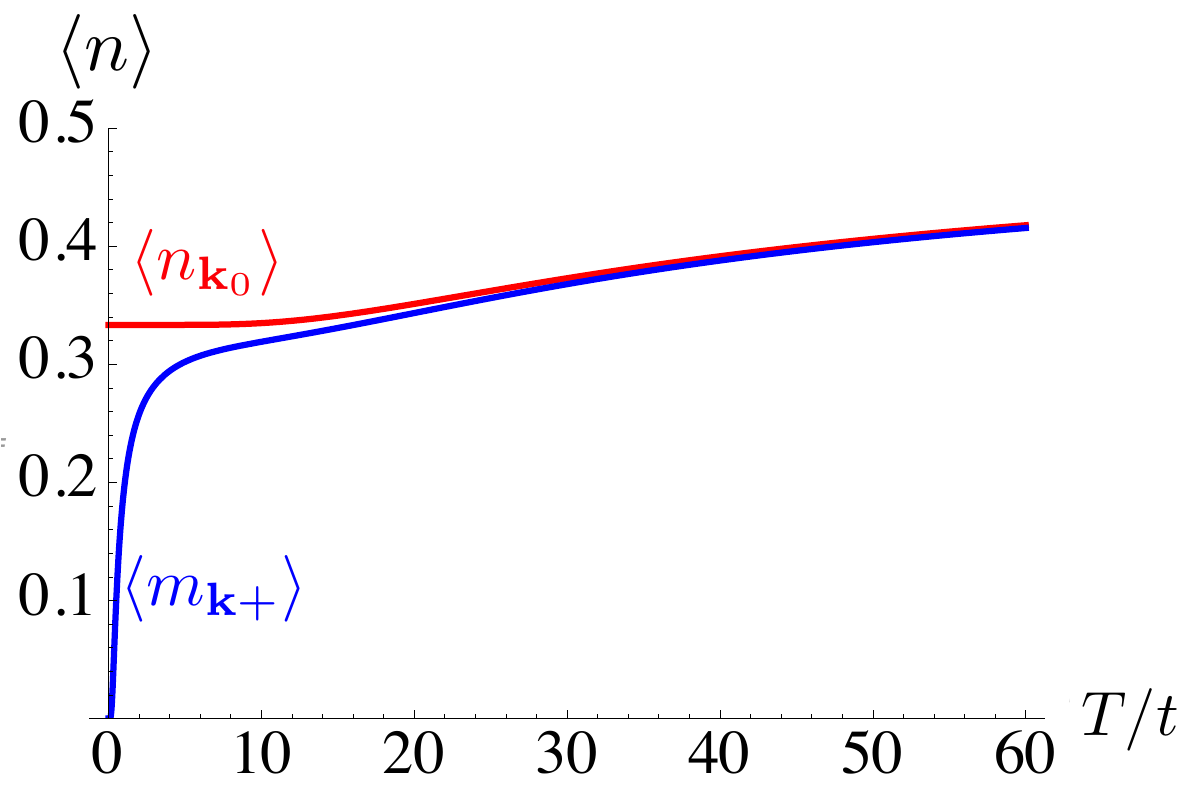}
    \caption{Fermionic occupation as a function of scaled temperature for $U/t=50,M=2,\mu=0$ , the non-Fermi liquid phase, for the Weyl momentum $\textbf{k}_0$ in red and a generic unoccupied (at $T=0$) momentum state $\textbf{k}=\textbf{k}_0+\delta\textbf{k}$ in blue. }
    \label{fig:finTemp}
 \end{figure}
Both facts are easily calculated from the occupation $\expval{m_{\textbf{k}_0 s}}$, which now has the special values of $\frac{1}{3}$ and $\frac{2}{3}$ at $\mu=0$ and $\mu=U$ respectively. The spectral function at the Weyl-Mott point then has the form
\begin{equation}
    \tilde{A}_s(\textbf{k}_0,\omega) = \left\{ \begin{array}{lll} \dfrac{4\pi}{3} \left( \delta(\omega)+\dfrac{1}{2}\delta(\omega-U)\right) & \qquad & \mu=0;\\
& & \\
    \dfrac{4\pi}{3} \left( \dfrac{1}{2}\delta(\omega+U)+\delta(\omega)\right) & & \mu=U. \end{array} \right.
\end{equation}
Such non-Fermi liquid behavior is robust up to a finite temperature as shown in Fig.~\ref{fig:finTemp}. The occupation at the Weyl momenta stays approximately constant up to a temperature of order of the interaction strength $U$, as can be seen analytically at $\mu=0$:
\begin{align}
    \expval{m_{\textbf{k}_0 s}}=\dfrac{1+e^{-\beta U}}{3+e^{-\beta U}}.
\end{align}
This is in contrast to the occupation at other momenta, which is heavily temperature-dependent at low temperatures.

What happens to the Green's function zero that was previously located at $\omega_0= U/2-\mu$ for $0<\mu<U$? Remarkably, its frequency changes discontinuously to $\omega_0 = \pm\frac{2U}{3}$ for $\mu=0,U$ respectively (see  Supplemental Material). We can thus identify the transition between the Weyl-Mott phase and the simple Fermi liquid as a critical point that displays non-Fermi liquid behavior and a discontinuous change in the frequency of the Green's function zero.

The fact that such critical behavior is found also for finite temperatures is consistent with previous theoretical and experimental settings such as Anderson impurity models \cite{TQPTAndersonimpurity}, strongly-correlated Weyl–Kondo semimetals without a quasiparticle description \cite{kirschbaum_how_2023,hu_topological_2022} and iron phthalocyanines on Au(111) \cite{zitko_iron_2021}. 
The fate of such NFL critical points under attractive instabilities, and the nature of any resulting superconducting state, will be analyzed in future work.
 
The topological charge of the Weyl-Mott points is the same as the non-interacting one \cite{armitage_weyl_2018}. It follows that the chiral anomaly coming from the change in the path-integral measure will still be there in our model; the interaction part only deals with the number of particles, with no phase-dependence, and thus does not contribute to this response. More concretely, since the chiral anomaly is perturbatively stable as long as translation symmetry is preserved, which is the case for \eqref{hWHK}, we conclude that Weyl-Mott points will also display an emergent chiral anomaly characterized by the following effective action (after integrating out the fermions):
\begin{align}
    S_\theta=\frac{1}{2 \pi} \frac{e^2}{h} \int \dd t\, \dd^3 \mathbf{r} \ \theta(\mathbf{r}, t) \ \mathbf{E} \cdot \mathbf{B},
\end{align}
 where $\theta(\mathbf{r}, t)$ is the axion term, given for non-interacting Weyl semimetals by $\theta(\mathbf{r}, t)=2\left(\mathbf{k}_0 \cdot \mathbf{r}-b_0 t\right)$, with $b_0$ a potential separation in energy between the Weyl points.
 
 As noted in ref.\ \cite{wang_chiral_2013} the $\theta$ field may be upgraded to a dynamical axion term in the presence of interactions. Although such an anomaly in the presence of strong interactions was also found in ref.\ \cite{meng_unpaired_2019}, here we extend this also to the NFL phase of Fig.~\ref{fig:phase} without any mean-field assumption and even away from the special $\mu=U/2$ point. In a picture where the fermions are retained, it is to be interpreted as coming from the topological charge of the Green's function zero (as opposed to the pole in the non-interacting setting). The spectrum, as shown before, is naturally spin-split; we interpret this as signaling that spin may play a role, perhaps indicating a spin density wave (SDW) rather than the previously assumed CDW in ref.\ \cite{burkovCDW_2020} and \cite{wang_chiral_2013}, consistent with other interacting Weyl models \cite{kundu_spin_2021}. 
 
\textit{Conclusion.} In this paper, we studied the phase diagram of a Hatsugai-Kohmoto-Weyl model that is a lattice completion of that presented in ref.\ \cite{morimoto_weyl_2016} as a function of the chemical potential. We found regions of insulating behavior which are not adiabatically connected to a simple non-interacting model, and where the Weyl points are gapped, and we showed that the extent of these regions is directly linked to the presence of an isolated topological Green's function zero. Such {\it Weyl-Mott points} have the same topological charge as the non-interacting Weyl points and are also associated to a chiral anomaly in this strongly-interacting setting. 

We also found a non-Fermi-liquid critical point between the Weyl-Mott insulator and a simple Fermi-liquid metal. This critical point is metallic, as it has spectral weight at zero frequency, but has a non-Fermi liquid behavior at the Weyl-Mott point. We showed that this remarkable property is robust to finite-temperature effects.  The onset of this non-Fermi-liquid behavior is accompanied by a discontinuous jump in the frequency of the Green's function zero. 

\textit{Acknowledgments.} We are grateful to Philip Phillips for very instructive comments and references. CH thanks the Max Planck Institute for the Physics of Complex Systems (MPI-PKS), where much of this work was carried out, for their hospitality.

\let\oldaddcontentsline\addcontentsline
\renewcommand{\addcontentsline}[3]{}
\bibliography{manuscript.bib}

\begin{thebibliography}{58}%
\makeatletter
\providecommand \@ifxundefined [1]{%
 \@ifx{#1\undefined}
}%
\providecommand \@ifnum [1]{%
 \ifnum #1\expandafter \@firstoftwo
 \else \expandafter \@secondoftwo
 \fi
}%
\providecommand \@ifx [1]{%
 \ifx #1\expandafter \@firstoftwo
 \else \expandafter \@secondoftwo
 \fi
}%
\providecommand \natexlab [1]{#1}%
\providecommand \enquote  [1]{``#1''}%
\providecommand \bibnamefont  [1]{#1}%
\providecommand \bibfnamefont [1]{#1}%
\providecommand \citenamefont [1]{#1}%
\providecommand \href@noop [0]{\@secondoftwo}%
\providecommand \href [0]{\begingroup \@sanitize@url \@href}%
\providecommand \@href[1]{\@@startlink{#1}\@@href}%
\providecommand \@@href[1]{\endgroup#1\@@endlink}%
\providecommand \@sanitize@url [0]{\catcode `\\12\catcode `\$12\catcode `\&12\catcode `\#12\catcode `\^12\catcode `\_12\catcode `\%12\relax}%
\providecommand \@@startlink[1]{}%
\providecommand \@@endlink[0]{}%
\providecommand \url  [0]{\begingroup\@sanitize@url \@url }%
\providecommand \@url [1]{\endgroup\@href {#1}{\urlprefix }}%
\providecommand \urlprefix  [0]{URL }%
\providecommand \Eprint [0]{\href }%
\providecommand \doibase [0]{https://doi.org/}%
\providecommand \selectlanguage [0]{\@gobble}%
\providecommand \bibinfo  [0]{\@secondoftwo}%
\providecommand \bibfield  [0]{\@secondoftwo}%
\providecommand \translation [1]{[#1]}%
\providecommand \BibitemOpen [0]{}%
\providecommand \bibitemStop [0]{}%
\providecommand \bibitemNoStop [0]{.\EOS\space}%
\providecommand \EOS [0]{\spacefactor3000\relax}%
\providecommand \BibitemShut  [1]{\csname bibitem#1\endcsname}%
\let\auto@bib@innerbib\@empty
\bibitem [{\citenamefont {Levin}\ and\ \citenamefont {Stern}(2009)}]{levin2009}%
  \BibitemOpen
  \bibfield  {author} {\bibinfo {author} {\bibfnamefont {M.}~\bibnamefont {Levin}}\ and\ \bibinfo {author} {\bibfnamefont {A.}~\bibnamefont {Stern}},\ }\bibfield  {title} {\bibinfo {title} {Fractional topological insulators},\ }\href {https://doi.org/10.1103/PhysRevLett.103.196803} {\bibfield  {journal} {\bibinfo  {journal} {Phys. Rev. Lett.}\ }\textbf {\bibinfo {volume} {103}},\ \bibinfo {pages} {196803} (\bibinfo {year} {2009})}\BibitemShut {NoStop}%
\bibitem [{\citenamefont {Swingle}(2010)}]{swingle2011}%
  \BibitemOpen
  \bibfield  {author} {\bibinfo {author} {\bibfnamefont {B.}~\bibnamefont {Swingle}},\ }\bibfield  {title} {\bibinfo {title} {Entanglement entropy and the fermi surface},\ }\href {https://doi.org/10.1103/PhysRevLett.105.050502} {\bibfield  {journal} {\bibinfo  {journal} {Phys. Rev. Lett.}\ }\textbf {\bibinfo {volume} {105}},\ \bibinfo {pages} {050502} (\bibinfo {year} {2010})}\BibitemShut {NoStop}%
\bibitem [{\citenamefont {Maciejko}\ \emph {et~al.}(2011)\citenamefont {Maciejko}, \citenamefont {Liu},\ and\ \citenamefont {Vishwanath}}]{maciejko2012}%
  \BibitemOpen
  \bibfield  {author} {\bibinfo {author} {\bibfnamefont {J.}~\bibnamefont {Maciejko}}, \bibinfo {author} {\bibfnamefont {Z.-X.}\ \bibnamefont {Liu}},\ and\ \bibinfo {author} {\bibfnamefont {A.}~\bibnamefont {Vishwanath}},\ }\bibfield  {title} {\bibinfo {title} {Topological order in a three-dimensional system of strongly interacting photons},\ }\href {https://doi.org/10.1103/PhysRevB.83.035111} {\bibfield  {journal} {\bibinfo  {journal} {Phys. Rev. B}\ }\textbf {\bibinfo {volume} {83}},\ \bibinfo {pages} {035111} (\bibinfo {year} {2011})}\BibitemShut {NoStop}%
\bibitem [{\citenamefont {Neupert}\ \emph {et~al.}(2011)\citenamefont {Neupert}, \citenamefont {Santos}, \citenamefont {Chamon},\ and\ \citenamefont {Mudry}}]{neupert2011}%
  \BibitemOpen
  \bibfield  {author} {\bibinfo {author} {\bibfnamefont {T.}~\bibnamefont {Neupert}}, \bibinfo {author} {\bibfnamefont {L.}~\bibnamefont {Santos}}, \bibinfo {author} {\bibfnamefont {C.}~\bibnamefont {Chamon}},\ and\ \bibinfo {author} {\bibfnamefont {C.}~\bibnamefont {Mudry}},\ }\bibfield  {title} {\bibinfo {title} {Fractional quantum hall states at zero magnetic field},\ }\href {https://doi.org/10.1103/PhysRevLett.106.236804} {\bibfield  {journal} {\bibinfo  {journal} {Phys. Rev. Lett.}\ }\textbf {\bibinfo {volume} {106}},\ \bibinfo {pages} {236804} (\bibinfo {year} {2011})}\BibitemShut {NoStop}%
\bibitem [{\citenamefont {Santos}\ \emph {et~al.}(2011)\citenamefont {Santos}, \citenamefont {Neupert}, \citenamefont {Ryu}, \citenamefont {Chamon},\ and\ \citenamefont {Mudry}}]{santos2011}%
  \BibitemOpen
  \bibfield  {author} {\bibinfo {author} {\bibfnamefont {L.}~\bibnamefont {Santos}}, \bibinfo {author} {\bibfnamefont {T.}~\bibnamefont {Neupert}}, \bibinfo {author} {\bibfnamefont {S.}~\bibnamefont {Ryu}}, \bibinfo {author} {\bibfnamefont {C.}~\bibnamefont {Chamon}},\ and\ \bibinfo {author} {\bibfnamefont {C.}~\bibnamefont {Mudry}},\ }\bibfield  {title} {\bibinfo {title} {Topological phase transitions in the spin-orbit coupled electron gas},\ }\href {https://doi.org/10.1103/PhysRevB.84.165138} {\bibfield  {journal} {\bibinfo  {journal} {Phys. Rev. B}\ }\textbf {\bibinfo {volume} {84}},\ \bibinfo {pages} {165138} (\bibinfo {year} {2011})}\BibitemShut {NoStop}%
\bibitem [{\citenamefont {Levin}\ and\ \citenamefont {Stern}(2012{\natexlab{a}})}]{levin2011exact}%
  \BibitemOpen
  \bibfield  {author} {\bibinfo {author} {\bibfnamefont {M.}~\bibnamefont {Levin}}\ and\ \bibinfo {author} {\bibfnamefont {A.}~\bibnamefont {Stern}},\ }\bibfield  {title} {\bibinfo {title} {Exactly soluble models for fractional topological insulators in two dimensions},\ }\href {https://doi.org/10.1103/PhysRevB.86.115131} {\bibfield  {journal} {\bibinfo  {journal} {Phys. Rev. B}\ }\textbf {\bibinfo {volume} {86}},\ \bibinfo {pages} {115131} (\bibinfo {year} {2012}{\natexlab{a}})}\BibitemShut {NoStop}%
\bibitem [{\citenamefont {Levin}\ and\ \citenamefont {Stern}(2012{\natexlab{b}})}]{levin2012}%
  \BibitemOpen
  \bibfield  {author} {\bibinfo {author} {\bibfnamefont {M.}~\bibnamefont {Levin}}\ and\ \bibinfo {author} {\bibfnamefont {A.}~\bibnamefont {Stern}},\ }\bibfield  {title} {\bibinfo {title} {Protected edge modes without symmetry},\ }\href {https://doi.org/10.1103/PhysRevB.86.115131} {\bibfield  {journal} {\bibinfo  {journal} {Phys. Rev. B}\ }\textbf {\bibinfo {volume} {86}},\ \bibinfo {pages} {115131} (\bibinfo {year} {2012}{\natexlab{b}})}\BibitemShut {NoStop}%
\bibitem [{\citenamefont {Chen}\ \emph {et~al.}(2012)\citenamefont {Chen}, \citenamefont {Gu}, \citenamefont {Liu},\ and\ \citenamefont {Wen}}]{chen2013}%
  \BibitemOpen
  \bibfield  {author} {\bibinfo {author} {\bibfnamefont {X.}~\bibnamefont {Chen}}, \bibinfo {author} {\bibfnamefont {Z.-C.}\ \bibnamefont {Gu}}, \bibinfo {author} {\bibfnamefont {Z.-X.}\ \bibnamefont {Liu}},\ and\ \bibinfo {author} {\bibfnamefont {X.-G.}\ \bibnamefont {Wen}},\ }\bibfield  {title} {\bibinfo {title} {Symmetry protected topological orders in interacting bosonic systems},\ }\href {https://www.science.org/doi/10.1126/science.1227224} {\bibfield  {journal} {\bibinfo  {journal} {Science}\ }\textbf {\bibinfo {volume} {338}},\ \bibinfo {pages} {1604} (\bibinfo {year} {2012})}\BibitemShut {NoStop}%
\bibitem [{\citenamefont {Gu}\ and\ \citenamefont {Wen}(2014)}]{gu2014}%
  \BibitemOpen
  \bibfield  {author} {\bibinfo {author} {\bibfnamefont {Z.-C.}\ \bibnamefont {Gu}}\ and\ \bibinfo {author} {\bibfnamefont {X.-G.}\ \bibnamefont {Wen}},\ }\bibfield  {title} {\bibinfo {title} {Symmetry-protected topological orders for interacting fermions: Fermionic topological nonlinear sigma models and a special group supercohomology theory},\ }\href {https://doi.org/10.1103/PhysRevB.90.115141} {\bibfield  {journal} {\bibinfo  {journal} {Phys. Rev. B}\ }\textbf {\bibinfo {volume} {90}},\ \bibinfo {pages} {115141} (\bibinfo {year} {2014})}\BibitemShut {NoStop}%
\bibitem [{\citenamefont {Vishwanath}\ and\ \citenamefont {Senthil}(2013)}]{vishwanath20133d}%
  \BibitemOpen
  \bibfield  {author} {\bibinfo {author} {\bibfnamefont {A.}~\bibnamefont {Vishwanath}}\ and\ \bibinfo {author} {\bibfnamefont {T.}~\bibnamefont {Senthil}},\ }\bibfield  {title} {\bibinfo {title} {Three-dimensional bosonic topological insulator: Symmetry-protected topological order, and the boundary degrees of freedom},\ }\href {https://doi.org/10.1103/PhysRevX.3.011016} {\bibfield  {journal} {\bibinfo  {journal} {Phys. Rev. X}\ }\textbf {\bibinfo {volume} {3}},\ \bibinfo {pages} {011016} (\bibinfo {year} {2013})}\BibitemShut {NoStop}%
\bibitem [{\citenamefont {Wang}\ and\ \citenamefont {Senthil}(2013)}]{wang2013bosonTI}%
  \BibitemOpen
  \bibfield  {author} {\bibinfo {author} {\bibfnamefont {C.}~\bibnamefont {Wang}}\ and\ \bibinfo {author} {\bibfnamefont {T.}~\bibnamefont {Senthil}},\ }\bibfield  {title} {\bibinfo {title} {Boson topological insulators: A window into highly entangled quantum phases},\ }\href {https://doi.org/10.1103/PhysRevB.87.235122} {\bibfield  {journal} {\bibinfo  {journal} {Phys. Rev. B}\ }\textbf {\bibinfo {volume} {87}},\ \bibinfo {pages} {235122} (\bibinfo {year} {2013})}\BibitemShut {NoStop}%
\bibitem [{\citenamefont {Burnell}\ \emph {et~al.}(2014)\citenamefont {Burnell}, \citenamefont {Chen}, \citenamefont {Fidkowski},\ and\ \citenamefont {Vishwanath}}]{burnell2014}%
  \BibitemOpen
  \bibfield  {author} {\bibinfo {author} {\bibfnamefont {F.~J.}\ \bibnamefont {Burnell}}, \bibinfo {author} {\bibfnamefont {X.}~\bibnamefont {Chen}}, \bibinfo {author} {\bibfnamefont {L.}~\bibnamefont {Fidkowski}},\ and\ \bibinfo {author} {\bibfnamefont {A.}~\bibnamefont {Vishwanath}},\ }\bibfield  {title} {\bibinfo {title} {Exactly solvable model of a three-dimensional symmetry-protected topological phase of bosons with surface topological order},\ }\href {https://doi.org/10.1103/PhysRevB.90.245122} {\bibfield  {journal} {\bibinfo  {journal} {Phys. Rev. B}\ }\textbf {\bibinfo {volume} {90}},\ \bibinfo {pages} {245122} (\bibinfo {year} {2014})}\BibitemShut {NoStop}%
\bibitem [{\citenamefont {Kapustin}(2014)}]{kapustin2014symmetry}%
  \BibitemOpen
  \bibfield  {author} {\bibinfo {author} {\bibfnamefont {A.}~\bibnamefont {Kapustin}},\ }\bibfield  {title} {\bibinfo {title} {Symmetry protected topological phases, anomalies, and cobordisms: Beyond group cohomology},\ }\href {https://arxiv.org/abs/1403.1467} {\bibfield  {journal} {\bibinfo  {journal} {arXiv preprint arXiv:1403.1467}\ } (\bibinfo {year} {2014})}\BibitemShut {NoStop}%
\bibitem [{\citenamefont {Fidkowski}\ \emph {et~al.}(2013)\citenamefont {Fidkowski}, \citenamefont {Chen},\ and\ \citenamefont {Vishwanath}}]{fidkowski2013}%
  \BibitemOpen
  \bibfield  {author} {\bibinfo {author} {\bibfnamefont {L.}~\bibnamefont {Fidkowski}}, \bibinfo {author} {\bibfnamefont {X.}~\bibnamefont {Chen}},\ and\ \bibinfo {author} {\bibfnamefont {A.}~\bibnamefont {Vishwanath}},\ }\bibfield  {title} {\bibinfo {title} {Non-abelian topological order on the surface of a 3d topological superconductor from an exactly solved model},\ }\href {https://doi.org/10.1103/PhysRevX.3.041016} {\bibfield  {journal} {\bibinfo  {journal} {Phys. Rev. X}\ }\textbf {\bibinfo {volume} {3}},\ \bibinfo {pages} {041016} (\bibinfo {year} {2013})}\BibitemShut {NoStop}%
\bibitem [{\citenamefont {Wang}\ and\ \citenamefont {Senthil}(2014)}]{wang2014}%
  \BibitemOpen
  \bibfield  {author} {\bibinfo {author} {\bibfnamefont {C.}~\bibnamefont {Wang}}\ and\ \bibinfo {author} {\bibfnamefont {T.}~\bibnamefont {Senthil}},\ }\bibfield  {title} {\bibinfo {title} {Interacting fermionic topological insulators/superconductors in three dimensions},\ }\href {https://doi.org/10.1103/PhysRevB.89.195124} {\bibfield  {journal} {\bibinfo  {journal} {Phys. Rev. B}\ }\textbf {\bibinfo {volume} {89}},\ \bibinfo {pages} {195124} (\bibinfo {year} {2014})}\BibitemShut {NoStop}%
\bibitem [{\citenamefont {Hsieh}\ \emph {et~al.}(2008)\citenamefont {Hsieh}, \citenamefont {Qian}, \citenamefont {Wray}, \citenamefont {Xia}, \citenamefont {Hor}, \citenamefont {Cava},\ and\ \citenamefont {Hasan}}]{3DTI-exp2}%
  \BibitemOpen
  \bibfield  {author} {\bibinfo {author} {\bibfnamefont {D.}~\bibnamefont {Hsieh}}, \bibinfo {author} {\bibfnamefont {D.}~\bibnamefont {Qian}}, \bibinfo {author} {\bibfnamefont {L.}~\bibnamefont {Wray}}, \bibinfo {author} {\bibfnamefont {Y.}~\bibnamefont {Xia}}, \bibinfo {author} {\bibfnamefont {Y.~S.}\ \bibnamefont {Hor}}, \bibinfo {author} {\bibfnamefont {R.~J.}\ \bibnamefont {Cava}},\ and\ \bibinfo {author} {\bibfnamefont {M.~Z.}\ \bibnamefont {Hasan}},\ }\bibfield  {title} {\bibinfo {title} {Topological insulators in three dimensions},\ }\href {https://doi.org/10.1038/nature06843} {\bibfield  {journal} {\bibinfo  {journal} {Nature}\ }\textbf {\bibinfo {volume} {452}},\ \bibinfo {pages} {970} (\bibinfo {year} {2008})}\BibitemShut {NoStop}%
\bibitem [{\citenamefont {Kane}\ and\ \citenamefont {Mele}(2005{\natexlab{a}})}]{kane2005z2}%
  \BibitemOpen
  \bibfield  {author} {\bibinfo {author} {\bibfnamefont {C.~L.}\ \bibnamefont {Kane}}\ and\ \bibinfo {author} {\bibfnamefont {E.~J.}\ \bibnamefont {Mele}},\ }\bibfield  {title} {\bibinfo {title} {Z$_2$ topological order and the quantum spin hall effect},\ }\href {https://doi.org/10.1103/PhysRevLett.95.146802} {\bibfield  {journal} {\bibinfo  {journal} {Phys. Rev. Lett.}\ }\textbf {\bibinfo {volume} {95}},\ \bibinfo {pages} {146802} (\bibinfo {year} {2005}{\natexlab{a}})}\BibitemShut {NoStop}%
\bibitem [{\citenamefont {Kane}\ and\ \citenamefont {Mele}(2005{\natexlab{b}})}]{kane2005quantum}%
  \BibitemOpen
  \bibfield  {author} {\bibinfo {author} {\bibfnamefont {C.~L.}\ \bibnamefont {Kane}}\ and\ \bibinfo {author} {\bibfnamefont {E.~J.}\ \bibnamefont {Mele}},\ }\bibfield  {title} {\bibinfo {title} {Quantum spin hall effect in graphene},\ }\href {https://doi.org/10.1103/PhysRevLett.95.226801} {\bibfield  {journal} {\bibinfo  {journal} {Phys. Rev. Lett.}\ }\textbf {\bibinfo {volume} {95}},\ \bibinfo {pages} {226801} (\bibinfo {year} {2005}{\natexlab{b}})}\BibitemShut {NoStop}%
\bibitem [{\citenamefont {Fu}\ \emph {et~al.}(2007)\citenamefont {Fu}, \citenamefont {Kane},\ and\ \citenamefont {Mele}}]{Fu-Kane-Mele}%
  \BibitemOpen
  \bibfield  {author} {\bibinfo {author} {\bibfnamefont {L.}~\bibnamefont {Fu}}, \bibinfo {author} {\bibfnamefont {C.~L.}\ \bibnamefont {Kane}},\ and\ \bibinfo {author} {\bibfnamefont {E.~J.}\ \bibnamefont {Mele}},\ }\bibfield  {title} {\bibinfo {title} {Topological insulators in three dimensions},\ }\href {https://doi.org/10.1103/PhysRevLett.98.106803} {\bibfield  {journal} {\bibinfo  {journal} {Phys. Rev. Lett.}\ }\textbf {\bibinfo {volume} {98}},\ \bibinfo {pages} {106803} (\bibinfo {year} {2007})}\BibitemShut {NoStop}%
\bibitem [{\citenamefont {Schnyder}\ \emph {et~al.}(2008)\citenamefont {Schnyder}, \citenamefont {Ryu}, \citenamefont {Furusaki},\ and\ \citenamefont {Ludwig}}]{ten-fold-way}%
  \BibitemOpen
  \bibfield  {author} {\bibinfo {author} {\bibfnamefont {A.~P.}\ \bibnamefont {Schnyder}}, \bibinfo {author} {\bibfnamefont {S.}~\bibnamefont {Ryu}}, \bibinfo {author} {\bibfnamefont {A.}~\bibnamefont {Furusaki}},\ and\ \bibinfo {author} {\bibfnamefont {A.~W.~W.}\ \bibnamefont {Ludwig}},\ }\bibfield  {title} {\bibinfo {title} {Classification of topological insulators and superconductors},\ }\href {https://doi.org/10.1103/PhysRevB.78.195125} {\bibfield  {journal} {\bibinfo  {journal} {Phys. Rev. B}\ }\textbf {\bibinfo {volume} {78}},\ \bibinfo {pages} {195125} (\bibinfo {year} {2008})}\BibitemShut {NoStop}%
\bibitem [{\citenamefont {Moore}\ and\ \citenamefont {Balents}(2007)}]{Moorehomotopy2007}%
  \BibitemOpen
  \bibfield  {author} {\bibinfo {author} {\bibfnamefont {J.~E.}\ \bibnamefont {Moore}}\ and\ \bibinfo {author} {\bibfnamefont {L.}~\bibnamefont {Balents}},\ }\bibfield  {title} {\bibinfo {title} {Topological invariants of time-reversal-invariant band structures},\ }\href {https://doi.org/10.1103/PhysRevB.75.121306} {\bibfield  {journal} {\bibinfo  {journal} {Phys. Rev. B}\ }\textbf {\bibinfo {volume} {75}},\ \bibinfo {pages} {121306} (\bibinfo {year} {2007})}\BibitemShut {NoStop}%
\bibitem [{\citenamefont {Haldane}(1983)}]{haldane1983}%
  \BibitemOpen
  \bibfield  {author} {\bibinfo {author} {\bibfnamefont {F.~D.~M.}\ \bibnamefont {Haldane}},\ }\bibfield  {title} {\bibinfo {title} {Continuum dynamics of the 1-d heisenberg antiferromagnet: Identification with the o(3) nonlinear sigma model},\ }\href {https://doi.org/https://doi.org/10.1016/0375-9601(83)90631-X} {\bibfield  {journal} {\bibinfo  {journal} {Physics Letters A}\ }\textbf {\bibinfo {volume} {93}},\ \bibinfo {pages} {464} (\bibinfo {year} {1983})}\BibitemShut {NoStop}%
\bibitem [{\citenamefont {Affleck}\ \emph {et~al.}(1988)\citenamefont {Affleck}, \citenamefont {Kennedy}, \citenamefont {Lieb},\ and\ \citenamefont {Tasaki}}]{affleck1988}%
  \BibitemOpen
  \bibfield  {author} {\bibinfo {author} {\bibfnamefont {I.}~\bibnamefont {Affleck}}, \bibinfo {author} {\bibfnamefont {T.}~\bibnamefont {Kennedy}}, \bibinfo {author} {\bibfnamefont {E.~H.}\ \bibnamefont {Lieb}},\ and\ \bibinfo {author} {\bibfnamefont {H.}~\bibnamefont {Tasaki}},\ }\bibfield  {title} {\bibinfo {title} {Valence bond ground states in isotropic quantum antiferromagnets},\ }\href {https://doi.org/10.1007/BF01218021} {\bibfield  {journal} {\bibinfo  {journal} {Communications in Mathematical Physics}\ }\textbf {\bibinfo {volume} {115}},\ \bibinfo {pages} {477} (\bibinfo {year} {1988})}\BibitemShut {NoStop}%
\bibitem [{\citenamefont {Chen}\ \emph {et~al.}(2011)\citenamefont {Chen}, \citenamefont {Gu},\ and\ \citenamefont {Wen}}]{chen2011_gap1Dspin}%
  \BibitemOpen
  \bibfield  {author} {\bibinfo {author} {\bibfnamefont {X.}~\bibnamefont {Chen}}, \bibinfo {author} {\bibfnamefont {Z.-C.}\ \bibnamefont {Gu}},\ and\ \bibinfo {author} {\bibfnamefont {X.-G.}\ \bibnamefont {Wen}},\ }\bibfield  {title} {\bibinfo {title} {Classification of gapped symmetric phases in one-dimensional spin systems},\ }\href {https://doi.org/10.1103/PhysRevB.83.035107} {\bibfield  {journal} {\bibinfo  {journal} {Phys. Rev. B}\ }\textbf {\bibinfo {volume} {83}},\ \bibinfo {pages} {035107} (\bibinfo {year} {2011})}\BibitemShut {NoStop}%
\bibitem [{\citenamefont {Senthil}(2015)}]{senthil2015}%
  \BibitemOpen
  \bibfield  {author} {\bibinfo {author} {\bibfnamefont {T.}~\bibnamefont {Senthil}},\ }\bibfield  {title} {\bibinfo {title} {Symmetry-protected topological phases of quantum matter},\ }\href {https://doi.org/10.1146/annurev-conmatphys-031214-014740} {\bibfield  {journal} {\bibinfo  {journal} {Annual Review of Condensed Matter Physics}\ }\textbf {\bibinfo {volume} {6}},\ \bibinfo {pages} {299} (\bibinfo {year} {2015})}\BibitemShut {NoStop}%
\bibitem [{\citenamefont {Verresen}\ \emph {et~al.}(2017)\citenamefont {Verresen}, \citenamefont {Moessner},\ and\ \citenamefont {Pollmann}}]{Frank-Ruben-1d}%
  \BibitemOpen
  \bibfield  {author} {\bibinfo {author} {\bibfnamefont {R.}~\bibnamefont {Verresen}}, \bibinfo {author} {\bibfnamefont {R.}~\bibnamefont {Moessner}},\ and\ \bibinfo {author} {\bibfnamefont {F.}~\bibnamefont {Pollmann}},\ }\bibfield  {title} {\bibinfo {title} {One-dimensional symmetry protected topological phases and their transitions},\ }\href {https://doi.org/10.1103/PhysRevB.96.165124} {\bibfield  {journal} {\bibinfo  {journal} {Phys. Rev. B}\ }\textbf {\bibinfo {volume} {96}},\ \bibinfo {pages} {165124} (\bibinfo {year} {2017})}\BibitemShut {NoStop}%
\bibitem [{\citenamefont {Verresen}\ \emph {et~al.}(2018)\citenamefont {Verresen}, \citenamefont {Jones},\ and\ \citenamefont {Pollmann}}]{Ruben-TopoCritQuant}%
  \BibitemOpen
  \bibfield  {author} {\bibinfo {author} {\bibfnamefont {R.}~\bibnamefont {Verresen}}, \bibinfo {author} {\bibfnamefont {N.~G.}\ \bibnamefont {Jones}},\ and\ \bibinfo {author} {\bibfnamefont {F.}~\bibnamefont {Pollmann}},\ }\bibfield  {title} {\bibinfo {title} {Topology and edge modes in quantum critical chains},\ }\href {https://doi.org/10.1103/PhysRevLett.120.057001} {\bibfield  {journal} {\bibinfo  {journal} {Phys. Rev. Lett.}\ }\textbf {\bibinfo {volume} {120}},\ \bibinfo {pages} {057001} (\bibinfo {year} {2018})}\BibitemShut {NoStop}%
\bibitem [{\citenamefont {Scaffidi}\ \emph {et~al.}(2017)\citenamefont {Scaffidi}, \citenamefont {Parker},\ and\ \citenamefont {Vasseur}}]{scaffidi2017}%
  \BibitemOpen
  \bibfield  {author} {\bibinfo {author} {\bibfnamefont {T.}~\bibnamefont {Scaffidi}}, \bibinfo {author} {\bibfnamefont {D.~E.}\ \bibnamefont {Parker}},\ and\ \bibinfo {author} {\bibfnamefont {R.}~\bibnamefont {Vasseur}},\ }\bibfield  {title} {\bibinfo {title} {Gapless symmetry-protected topological order},\ }\href {https://doi.org/10.1103/PhysRevX.7.041048} {\bibfield  {journal} {\bibinfo  {journal} {Phys. Rev. X}\ }\textbf {\bibinfo {volume} {7}},\ \bibinfo {pages} {041048} (\bibinfo {year} {2017})}\BibitemShut {NoStop}%
\bibitem [{\citenamefont {Thorngren}\ \emph {et~al.}(2021)\citenamefont {Thorngren}, \citenamefont {Vishwanath},\ and\ \citenamefont {Verresen}}]{Thorngren-PRB-2021}%
  \BibitemOpen
  \bibfield  {author} {\bibinfo {author} {\bibfnamefont {R.}~\bibnamefont {Thorngren}}, \bibinfo {author} {\bibfnamefont {A.}~\bibnamefont {Vishwanath}},\ and\ \bibinfo {author} {\bibfnamefont {R.}~\bibnamefont {Verresen}},\ }\bibfield  {title} {\bibinfo {title} {Intrinsically gapless topological phases},\ }\href {https://doi.org/10.1103/PhysRevB.104.075132} {\bibfield  {journal} {\bibinfo  {journal} {Phys. Rev. B}\ }\textbf {\bibinfo {volume} {104}},\ \bibinfo {pages} {075132} (\bibinfo {year} {2021})}\BibitemShut {NoStop}%
\bibitem [{\citenamefont {Verresen}\ \emph {et~al.}(2024)\citenamefont {Verresen}, \citenamefont {Borla}, \citenamefont {Vishwanath}, \citenamefont {Moroz},\ and\ \citenamefont {Thorngren}}]{verresen2024higgs}%
  \BibitemOpen
  \bibfield  {author} {\bibinfo {author} {\bibfnamefont {R.}~\bibnamefont {Verresen}}, \bibinfo {author} {\bibfnamefont {U.}~\bibnamefont {Borla}}, \bibinfo {author} {\bibfnamefont {A.}~\bibnamefont {Vishwanath}}, \bibinfo {author} {\bibfnamefont {S.}~\bibnamefont {Moroz}},\ and\ \bibinfo {author} {\bibfnamefont {R.}~\bibnamefont {Thorngren}},\ }\href {https://arxiv.org/abs/2211.01376} {\bibinfo {title} {Higgs condensates are symmetry-protected topological phases: I. discrete symmetries}} (\bibinfo {year} {2024}),\ \Eprint {https://arxiv.org/abs/2211.01376} {arXiv:2211.01376 [cond-mat.str-el]} \BibitemShut {NoStop}%
\bibitem [{\citenamefont {Chung}\ \emph {et~al.}(2024)\citenamefont {Chung}, \citenamefont {Flores-Calderón}, \citenamefont {Torres}, \citenamefont {Ribeiro}, \citenamefont {Moroz},\ and\ \citenamefont {McClarty}}]{chung2024higgs}%
  \BibitemOpen
  \bibfield  {author} {\bibinfo {author} {\bibfnamefont {K.~T.~K.}\ \bibnamefont {Chung}}, \bibinfo {author} {\bibfnamefont {R.}~\bibnamefont {Flores-Calderón}}, \bibinfo {author} {\bibfnamefont {R.~C.}\ \bibnamefont {Torres}}, \bibinfo {author} {\bibfnamefont {P.}~\bibnamefont {Ribeiro}}, \bibinfo {author} {\bibfnamefont {S.}~\bibnamefont {Moroz}},\ and\ \bibinfo {author} {\bibfnamefont {P.}~\bibnamefont {McClarty}},\ }\href {https://arxiv.org/abs/2404.17001} {\bibinfo {title} {Higgs phases and boundary criticality}} (\bibinfo {year} {2024}),\ \Eprint {https://arxiv.org/abs/2404.17001} {arXiv:2404.17001 [cond-mat.str-el]} \BibitemShut {NoStop}%
\bibitem [{\citenamefont {Peralta~Gavensky}\ \emph {et~al.}(2023)\citenamefont {Peralta~Gavensky}, \citenamefont {Sachdev},\ and\ \citenamefont {Goldman}}]{sachdev_chern_2023}%
  \BibitemOpen
  \bibfield  {author} {\bibinfo {author} {\bibfnamefont {L.}~\bibnamefont {Peralta~Gavensky}}, \bibinfo {author} {\bibfnamefont {S.}~\bibnamefont {Sachdev}},\ and\ \bibinfo {author} {\bibfnamefont {N.}~\bibnamefont {Goldman}},\ }\bibfield  {title} {\bibinfo {title} {Connecting the {Many}-{Body} {Chern} {Number} to {Luttinger}’s {Theorem} through {Streda}’s {Formula}},\ }\href {https://doi.org/10.1103/PhysRevLett.131.236601} {\bibfield  {journal} {\bibinfo  {journal} {Physical Review Letters}\ }\textbf {\bibinfo {volume} {131}},\ \bibinfo {pages} {236601} (\bibinfo {year} {2023})}\BibitemShut {NoStop}%
\bibitem [{\citenamefont {Mai}\ \emph {et~al.}(2023)\citenamefont {Mai}, \citenamefont {Feldman},\ and\ \citenamefont {Phillips}}]{mai2023}%
  \BibitemOpen
  \bibfield  {author} {\bibinfo {author} {\bibfnamefont {P.}~\bibnamefont {Mai}}, \bibinfo {author} {\bibfnamefont {B.~E.}\ \bibnamefont {Feldman}},\ and\ \bibinfo {author} {\bibfnamefont {P.~W.}\ \bibnamefont {Phillips}},\ }\bibfield  {title} {\bibinfo {title} {Topological mott insulator at quarter filling in the interacting haldane model},\ }\href {https://doi.org/10.1103/PhysRevResearch.5.013162} {\bibfield  {journal} {\bibinfo  {journal} {Phys. Rev. Res.}\ }\textbf {\bibinfo {volume} {5}},\ \bibinfo {pages} {013162} (\bibinfo {year} {2023})}\BibitemShut {NoStop}%
\bibitem [{\citenamefont {Bollmann}\ \emph {et~al.}(2023)\citenamefont {Bollmann}, \citenamefont {Setty}, \citenamefont {Seifert},\ and\ \citenamefont {König}}]{bollmann_topological_2023}%
  \BibitemOpen
  \bibfield  {author} {\bibinfo {author} {\bibfnamefont {S.}~\bibnamefont {Bollmann}}, \bibinfo {author} {\bibfnamefont {C.}~\bibnamefont {Setty}}, \bibinfo {author} {\bibfnamefont {U.~F.~P.}\ \bibnamefont {Seifert}},\ and\ \bibinfo {author} {\bibfnamefont {E.~J.}\ \bibnamefont {König}},\ }\href {http://arxiv.org/abs/2312.14926} {{\selectlanguage {en}\bibinfo {title} {Topological {Green}'s function zeros in an exactly solved model and beyond}}} (\bibinfo {year} {2023}),\ \bibinfo {note} {arXiv:2312.14926 [cond-mat]}\BibitemShut {NoStop}%
\bibitem [{\citenamefont {Wagner}\ \emph {et~al.}(2023)\citenamefont {Wagner}, \citenamefont {Crippa}, \citenamefont {Amaricci}, \citenamefont {Hansmann}, \citenamefont {Klett}, \citenamefont {König}, \citenamefont {Schäfer}, \citenamefont {Di~Sante}, \citenamefont {Cano}, \citenamefont {Millis}, \citenamefont {Georges},\ and\ \citenamefont {Sangiovanni}}]{wagner_mott_2023}%
  \BibitemOpen
  \bibfield  {author} {\bibinfo {author} {\bibfnamefont {N.}~\bibnamefont {Wagner}}, \bibinfo {author} {\bibfnamefont {L.}~\bibnamefont {Crippa}}, \bibinfo {author} {\bibfnamefont {A.}~\bibnamefont {Amaricci}}, \bibinfo {author} {\bibfnamefont {P.}~\bibnamefont {Hansmann}}, \bibinfo {author} {\bibfnamefont {M.}~\bibnamefont {Klett}}, \bibinfo {author} {\bibfnamefont {E.}~\bibnamefont {König}}, \bibinfo {author} {\bibfnamefont {T.}~\bibnamefont {Schäfer}}, \bibinfo {author} {\bibfnamefont {D.}~\bibnamefont {Di~Sante}}, \bibinfo {author} {\bibfnamefont {J.}~\bibnamefont {Cano}}, \bibinfo {author} {\bibfnamefont {A.}~\bibnamefont {Millis}}, \bibinfo {author} {\bibfnamefont {A.}~\bibnamefont {Georges}},\ and\ \bibinfo {author} {\bibfnamefont {G.}~\bibnamefont {Sangiovanni}},\ }\bibfield  {title} {{\selectlanguage {en}\bibinfo {title} {Mott insulators with boundary zeros}},\ }\href {https://doi.org/10.1038/s41467-023-42773-7} {\bibfield  {journal} {\bibinfo  {journal} {Nature Communications}\ }\textbf {\bibinfo
  {volume} {14}},\ \bibinfo {pages} {7531} (\bibinfo {year} {2023})},\ \bibinfo {note} {arXiv:2301.05588 [cond-mat]}\BibitemShut {NoStop}%
\bibitem [{\citenamefont {Setty}\ \emph {et~al.}(2023{\natexlab{a}})\citenamefont {Setty}, \citenamefont {Xie}, \citenamefont {Sur}, \citenamefont {Chen}, \citenamefont {Vergniory},\ and\ \citenamefont {Si}}]{setty_electronic_2023}%
  \BibitemOpen
  \bibfield  {author} {\bibinfo {author} {\bibfnamefont {C.}~\bibnamefont {Setty}}, \bibinfo {author} {\bibfnamefont {F.}~\bibnamefont {Xie}}, \bibinfo {author} {\bibfnamefont {S.}~\bibnamefont {Sur}}, \bibinfo {author} {\bibfnamefont {L.}~\bibnamefont {Chen}}, \bibinfo {author} {\bibfnamefont {M.~G.}\ \bibnamefont {Vergniory}},\ and\ \bibinfo {author} {\bibfnamefont {Q.}~\bibnamefont {Si}},\ }\href {http://arxiv.org/abs/2309.14340} {{\selectlanguage {en}\bibinfo {title} {Electronic properties, correlated topology and {Green}'s function zeros}}} (\bibinfo {year} {2023}{\natexlab{a}}),\ \bibinfo {note} {arXiv:2309.14340 [cond-mat]}\BibitemShut {NoStop}%
\bibitem [{\citenamefont {Setty}\ \emph {et~al.}(2023{\natexlab{b}})\citenamefont {Setty}, \citenamefont {Sur}, \citenamefont {Chen}, \citenamefont {Xie}, \citenamefont {Hu}, \citenamefont {Paschen}, \citenamefont {Cano},\ and\ \citenamefont {Si}}]{setty_symmetry_2023}%
  \BibitemOpen
  \bibfield  {author} {\bibinfo {author} {\bibfnamefont {C.}~\bibnamefont {Setty}}, \bibinfo {author} {\bibfnamefont {S.}~\bibnamefont {Sur}}, \bibinfo {author} {\bibfnamefont {L.}~\bibnamefont {Chen}}, \bibinfo {author} {\bibfnamefont {F.}~\bibnamefont {Xie}}, \bibinfo {author} {\bibfnamefont {H.}~\bibnamefont {Hu}}, \bibinfo {author} {\bibfnamefont {S.}~\bibnamefont {Paschen}}, \bibinfo {author} {\bibfnamefont {J.}~\bibnamefont {Cano}},\ and\ \bibinfo {author} {\bibfnamefont {Q.}~\bibnamefont {Si}},\ }\href {http://arxiv.org/abs/2301.13870} {{\selectlanguage {en}\bibinfo {title} {Symmetry constraints and spectral crossing in a {Mott} insulator with {Green}'s function zeros}}} (\bibinfo {year} {2023}{\natexlab{b}}),\ \bibinfo {note} {arXiv:2301.13870 [cond-mat]}\BibitemShut {NoStop}%
\bibitem [{\citenamefont {Blason}\ and\ \citenamefont {Fabrizio}(2023)}]{blason_unified_2023}%
  \BibitemOpen
  \bibfield  {author} {\bibinfo {author} {\bibfnamefont {A.}~\bibnamefont {Blason}}\ and\ \bibinfo {author} {\bibfnamefont {M.}~\bibnamefont {Fabrizio}},\ }\bibfield  {title} {{\selectlanguage {en}\bibinfo {title} {Unified role of {Green}'s function poles and zeros in correlated topological insulators}},\ }\href {https://doi.org/10.1103/PhysRevB.108.125115} {\bibfield  {journal} {\bibinfo  {journal} {Physical Review B}\ }\textbf {\bibinfo {volume} {108}},\ \bibinfo {pages} {125115} (\bibinfo {year} {2023})}\BibitemShut {NoStop}%
\bibitem [{\citenamefont {Hatsugai}\ and\ \citenamefont {Kohmoto}(1992)}]{hatsugai1992}%
  \BibitemOpen
  \bibfield  {author} {\bibinfo {author} {\bibfnamefont {Y.}~\bibnamefont {Hatsugai}}\ and\ \bibinfo {author} {\bibfnamefont {M.}~\bibnamefont {Kohmoto}},\ }\bibfield  {title} {\bibinfo {title} {Exactly solvable model of correlated lattice electrons in any dimensions},\ }\href {https://doi.org/10.1143/JPSJ.61.2056} {\bibfield  {journal} {\bibinfo  {journal} {Journal of the Physical Society of Japan}\ }\textbf {\bibinfo {volume} {61}},\ \bibinfo {pages} {2056} (\bibinfo {year} {1992})},\ \Eprint {https://arxiv.org/abs/https://doi.org/10.1143/JPSJ.61.2056} {https://doi.org/10.1143/JPSJ.61.2056} \BibitemShut {NoStop}%
\bibitem [{\citenamefont {Phillips}\ \emph {et~al.}(2020)\citenamefont {Phillips}, \citenamefont {Yeo},\ and\ \citenamefont {Huang}}]{phillips_exact_2020}%
  \BibitemOpen
  \bibfield  {author} {\bibinfo {author} {\bibfnamefont {P.~W.}\ \bibnamefont {Phillips}}, \bibinfo {author} {\bibfnamefont {L.}~\bibnamefont {Yeo}},\ and\ \bibinfo {author} {\bibfnamefont {E.~W.}\ \bibnamefont {Huang}},\ }\bibfield  {title} {{\selectlanguage {en}\bibinfo {title} {Exact theory for superconductivity in a doped {Mott} insulator}},\ }\href {https://doi.org/10.1038/s41567-020-0988-4} {\bibfield  {journal} {\bibinfo  {journal} {Nature Physics}\ }\textbf {\bibinfo {volume} {16}},\ \bibinfo {pages} {1175} (\bibinfo {year} {2020})}\BibitemShut {NoStop}%
\bibitem [{\citenamefont {Zhao}\ \emph {et~al.}(2023)\citenamefont {Zhao}, \citenamefont {La~Nave},\ and\ \citenamefont {Phillips}}]{zhao_proof_2023}%
  \BibitemOpen
  \bibfield  {author} {\bibinfo {author} {\bibfnamefont {J.}~\bibnamefont {Zhao}}, \bibinfo {author} {\bibfnamefont {G.}~\bibnamefont {La~Nave}},\ and\ \bibinfo {author} {\bibfnamefont {P.}~\bibnamefont {Phillips}},\ }\href {http://arxiv.org/abs/2304.04787} {{\selectlanguage {en}\bibinfo {title} {Proof of a {Stable} {Fixed} {Point} for {Strongly} {Correlated} {Electron} {Matter}}}} (\bibinfo {year} {2023}),\ \bibinfo {note} {arXiv:2304.04787 [cond-mat]}\BibitemShut {NoStop}%
\bibitem [{\citenamefont {Huang}\ \emph {et~al.}(2022)\citenamefont {Huang}, \citenamefont {Nave},\ and\ \citenamefont {Phillips}}]{huang_discrete_2022}%
  \BibitemOpen
  \bibfield  {author} {\bibinfo {author} {\bibfnamefont {E.~W.}\ \bibnamefont {Huang}}, \bibinfo {author} {\bibfnamefont {G.~L.}\ \bibnamefont {Nave}},\ and\ \bibinfo {author} {\bibfnamefont {P.~W.}\ \bibnamefont {Phillips}},\ }\bibfield  {title} {{\selectlanguage {en}\bibinfo {title} {Discrete symmetry breaking defines the {Mott} quartic fixed point}},\ }\href {https://doi.org/10.1038/s41567-022-01529-8} {\bibfield  {journal} {\bibinfo  {journal} {Nature Physics}\ }\textbf {\bibinfo {volume} {18}},\ \bibinfo {pages} {511} (\bibinfo {year} {2022})}\BibitemShut {NoStop}%
\bibitem [{\citenamefont {Armitage}\ \emph {et~al.}(2018)\citenamefont {Armitage}, \citenamefont {Mele},\ and\ \citenamefont {Vishwanath}}]{armitage_weyl_2018}%
  \BibitemOpen
  \bibfield  {author} {\bibinfo {author} {\bibfnamefont {N.}~\bibnamefont {Armitage}}, \bibinfo {author} {\bibfnamefont {E.}~\bibnamefont {Mele}},\ and\ \bibinfo {author} {\bibfnamefont {A.}~\bibnamefont {Vishwanath}},\ }\bibfield  {title} {{\selectlanguage {en}\bibinfo {title} {Weyl and {Dirac} semimetals in three-dimensional solids}},\ }\href {https://doi.org/10.1103/RevModPhys.90.015001} {\bibfield  {journal} {\bibinfo  {journal} {Reviews of Modern Physics}\ }\textbf {\bibinfo {volume} {90}},\ \bibinfo {pages} {015001} (\bibinfo {year} {2018})}\BibitemShut {NoStop}%
\bibitem [{\citenamefont {Wang}\ and\ \citenamefont {Zhang}(2013)}]{wang_chiral_2013}%
  \BibitemOpen
  \bibfield  {author} {\bibinfo {author} {\bibfnamefont {Z.}~\bibnamefont {Wang}}\ and\ \bibinfo {author} {\bibfnamefont {S.-C.}\ \bibnamefont {Zhang}},\ }\bibfield  {title} {{\selectlanguage {en}\bibinfo {title} {Chiral anomaly, charge density waves, and axion strings from {Weyl} semimetals}},\ }\href {https://doi.org/10.1103/PhysRevB.87.161107} {\bibfield  {journal} {\bibinfo  {journal} {Physical Review B}\ }\textbf {\bibinfo {volume} {87}},\ \bibinfo {pages} {161107} (\bibinfo {year} {2013})}\BibitemShut {NoStop}%
\bibitem [{\citenamefont {Zyuzin}\ and\ \citenamefont {Burkov}(2012)}]{zyuzin_topological_2012}%
  \BibitemOpen
  \bibfield  {author} {\bibinfo {author} {\bibfnamefont {A.~A.}\ \bibnamefont {Zyuzin}}\ and\ \bibinfo {author} {\bibfnamefont {A.~A.}\ \bibnamefont {Burkov}},\ }\bibfield  {title} {\bibinfo {title} {Topological response in {Weyl} semimetals and the chiral anomaly},\ }\href {https://doi.org/10.1103/PhysRevB.86.115133} {\bibfield  {journal} {\bibinfo  {journal} {Physical Review B}\ }\textbf {\bibinfo {volume} {86}},\ \bibinfo {pages} {115133} (\bibinfo {year} {2012})}\BibitemShut {NoStop}%
\bibitem [{\citenamefont {Sehayek}\ \emph {et~al.}(2020)\citenamefont {Sehayek}, \citenamefont {Thakurathi},\ and\ \citenamefont {Burkov}}]{burkovCDW_2020}%
  \BibitemOpen
  \bibfield  {author} {\bibinfo {author} {\bibfnamefont {D.}~\bibnamefont {Sehayek}}, \bibinfo {author} {\bibfnamefont {M.}~\bibnamefont {Thakurathi}},\ and\ \bibinfo {author} {\bibfnamefont {A.~A.}\ \bibnamefont {Burkov}},\ }\bibfield  {title} {{\selectlanguage {en}\bibinfo {title} {Charge density waves in {Weyl} semimetals}},\ }\href {https://doi.org/10.1103/PhysRevB.102.115159} {\bibfield  {journal} {\bibinfo  {journal} {Physical Review B}\ }\textbf {\bibinfo {volume} {102}},\ \bibinfo {pages} {115159} (\bibinfo {year} {2020})}\BibitemShut {NoStop}%
\bibitem [{\citenamefont {Crippa}\ \emph {et~al.}(2020)\citenamefont {Crippa}, \citenamefont {Amaricci}, \citenamefont {Wagner}, \citenamefont {Sangiovanni}, \citenamefont {Budich},\ and\ \citenamefont {Capone}}]{crippa_nonlocal_2020}%
  \BibitemOpen
  \bibfield  {author} {\bibinfo {author} {\bibfnamefont {L.}~\bibnamefont {Crippa}}, \bibinfo {author} {\bibfnamefont {A.}~\bibnamefont {Amaricci}}, \bibinfo {author} {\bibfnamefont {N.}~\bibnamefont {Wagner}}, \bibinfo {author} {\bibfnamefont {G.}~\bibnamefont {Sangiovanni}}, \bibinfo {author} {\bibfnamefont {J.~C.}\ \bibnamefont {Budich}},\ and\ \bibinfo {author} {\bibfnamefont {M.}~\bibnamefont {Capone}},\ }\bibfield  {title} {{\selectlanguage {en}\bibinfo {title} {Nonlocal annihilation of {Weyl} fermions in correlated systems}},\ }\href {https://doi.org/10.1103/PhysRevResearch.2.012023} {\bibfield  {journal} {\bibinfo  {journal} {Physical Review Research}\ }\textbf {\bibinfo {volume} {2}},\ \bibinfo {pages} {012023} (\bibinfo {year} {2020})}\BibitemShut {NoStop}%
\bibitem [{\citenamefont {Bobrow}\ \emph {et~al.}(2020)\citenamefont {Bobrow}, \citenamefont {Sun},\ and\ \citenamefont {Li}}]{bobrow_monopole_2020}%
  \BibitemOpen
  \bibfield  {author} {\bibinfo {author} {\bibfnamefont {E.}~\bibnamefont {Bobrow}}, \bibinfo {author} {\bibfnamefont {C.}~\bibnamefont {Sun}},\ and\ \bibinfo {author} {\bibfnamefont {Y.}~\bibnamefont {Li}},\ }\bibfield  {title} {{\selectlanguage {en}\bibinfo {title} {Monopole charge density wave states in {Weyl} semimetals}},\ }\href {https://doi.org/10.1103/PhysRevResearch.2.012078} {\bibfield  {journal} {\bibinfo  {journal} {Physical Review Research}\ }\textbf {\bibinfo {volume} {2}},\ \bibinfo {pages} {012078} (\bibinfo {year} {2020})}\BibitemShut {NoStop}%
\bibitem [{\citenamefont {Shi}\ \emph {et~al.}(2021)\citenamefont {Shi}, \citenamefont {Wieder}, \citenamefont {Meyerheim}, \citenamefont {Sun}, \citenamefont {Zhang}, \citenamefont {Li}, \citenamefont {Shen}, \citenamefont {Qi}, \citenamefont {Yang}, \citenamefont {Jena}, \citenamefont {Werner}, \citenamefont {Koepernik}, \citenamefont {Parkin}, \citenamefont {Chen}, \citenamefont {Felser}, \citenamefont {Bernevig},\ and\ \citenamefont {Wang}}]{shi_charge-density-wave_2021}%
  \BibitemOpen
  \bibfield  {author} {\bibinfo {author} {\bibfnamefont {W.}~\bibnamefont {Shi}}, \bibinfo {author} {\bibfnamefont {B.~J.}\ \bibnamefont {Wieder}}, \bibinfo {author} {\bibfnamefont {H.~L.}\ \bibnamefont {Meyerheim}}, \bibinfo {author} {\bibfnamefont {Y.}~\bibnamefont {Sun}}, \bibinfo {author} {\bibfnamefont {Y.}~\bibnamefont {Zhang}}, \bibinfo {author} {\bibfnamefont {Y.}~\bibnamefont {Li}}, \bibinfo {author} {\bibfnamefont {L.}~\bibnamefont {Shen}}, \bibinfo {author} {\bibfnamefont {Y.}~\bibnamefont {Qi}}, \bibinfo {author} {\bibfnamefont {L.}~\bibnamefont {Yang}}, \bibinfo {author} {\bibfnamefont {J.}~\bibnamefont {Jena}}, \bibinfo {author} {\bibfnamefont {P.}~\bibnamefont {Werner}}, \bibinfo {author} {\bibfnamefont {K.}~\bibnamefont {Koepernik}}, \bibinfo {author} {\bibfnamefont {S.}~\bibnamefont {Parkin}}, \bibinfo {author} {\bibfnamefont {Y.}~\bibnamefont {Chen}}, \bibinfo {author} {\bibfnamefont {C.}~\bibnamefont {Felser}}, \bibinfo {author} {\bibfnamefont {B.~A.}\ \bibnamefont {Bernevig}},\ and\
  \bibinfo {author} {\bibfnamefont {Z.}~\bibnamefont {Wang}},\ }\bibfield  {title} {{\selectlanguage {en}\bibinfo {title} {A charge-density-wave topological semimetal}},\ }\href {https://doi.org/10.1038/s41567-020-01104-z} {\bibfield  {journal} {\bibinfo  {journal} {Nature Physics}\ }\textbf {\bibinfo {volume} {17}},\ \bibinfo {pages} {381} (\bibinfo {year} {2021})}\BibitemShut {NoStop}%
\bibitem [{\citenamefont {Kirschbaum}\ \emph {et~al.}(2023)\citenamefont {Kirschbaum}, \citenamefont {Lužnik}, \citenamefont {Roy},\ and\ \citenamefont {Paschen}}]{kirschbaum_how_2023}%
  \BibitemOpen
  \bibfield  {author} {\bibinfo {author} {\bibfnamefont {D.~M.}\ \bibnamefont {Kirschbaum}}, \bibinfo {author} {\bibfnamefont {M.}~\bibnamefont {Lužnik}}, \bibinfo {author} {\bibfnamefont {G.~L.}\ \bibnamefont {Roy}},\ and\ \bibinfo {author} {\bibfnamefont {S.}~\bibnamefont {Paschen}},\ }\href {http://arxiv.org/abs/2308.11318} {{\selectlanguage {en}\bibinfo {title} {How to identify and characterize strongly correlated topological semimetals}}} (\bibinfo {year} {2023}),\ \bibinfo {note} {arXiv:2308.11318 [cond-mat]}\BibitemShut {NoStop}%
\bibitem [{\citenamefont {Hu}\ \emph {et~al.}(2022)\citenamefont {Hu}, \citenamefont {Chen}, \citenamefont {Setty}, \citenamefont {Garcia-Diez}, \citenamefont {Grefe}, \citenamefont {Prokofiev}, \citenamefont {Kirchner}, \citenamefont {Vergniory}, \citenamefont {Paschen}, \citenamefont {Cano},\ and\ \citenamefont {Si}}]{hu_topological_2022}%
  \BibitemOpen
  \bibfield  {author} {\bibinfo {author} {\bibfnamefont {H.}~\bibnamefont {Hu}}, \bibinfo {author} {\bibfnamefont {L.}~\bibnamefont {Chen}}, \bibinfo {author} {\bibfnamefont {C.}~\bibnamefont {Setty}}, \bibinfo {author} {\bibfnamefont {M.}~\bibnamefont {Garcia-Diez}}, \bibinfo {author} {\bibfnamefont {S.~E.}\ \bibnamefont {Grefe}}, \bibinfo {author} {\bibfnamefont {A.}~\bibnamefont {Prokofiev}}, \bibinfo {author} {\bibfnamefont {S.}~\bibnamefont {Kirchner}}, \bibinfo {author} {\bibfnamefont {M.~G.}\ \bibnamefont {Vergniory}}, \bibinfo {author} {\bibfnamefont {S.}~\bibnamefont {Paschen}}, \bibinfo {author} {\bibfnamefont {J.}~\bibnamefont {Cano}},\ and\ \bibinfo {author} {\bibfnamefont {Q.}~\bibnamefont {Si}},\ }\href {http://arxiv.org/abs/2110.06182} {{\selectlanguage {en}\bibinfo {title} {Topological semimetals without quasiparticles}}} (\bibinfo {year} {2022}),\ \bibinfo {note} {arXiv:2110.06182 [cond-mat]}\BibitemShut {NoStop}%
\bibitem [{\citenamefont {Morimoto}\ and\ \citenamefont {Nagaosa}(2016)}]{morimoto_weyl_2016}%
  \BibitemOpen
  \bibfield  {author} {\bibinfo {author} {\bibfnamefont {T.}~\bibnamefont {Morimoto}}\ and\ \bibinfo {author} {\bibfnamefont {N.}~\bibnamefont {Nagaosa}},\ }\bibfield  {title} {{\selectlanguage {en}\bibinfo {title} {Weyl {Mott} {Insulator}}},\ }\href {https://doi.org/10.1038/srep19853} {\bibfield  {journal} {\bibinfo  {journal} {Scientific Reports}\ }\textbf {\bibinfo {volume} {6}},\ \bibinfo {pages} {19853} (\bibinfo {year} {2016})}\BibitemShut {NoStop}%
\bibitem [{\citenamefont {Skolimowski}\ and\ \citenamefont {Fabrizio}(2022)}]{skolimowski_luttingers_2022}%
  \BibitemOpen
  \bibfield  {author} {\bibinfo {author} {\bibfnamefont {J.}~\bibnamefont {Skolimowski}}\ and\ \bibinfo {author} {\bibfnamefont {M.}~\bibnamefont {Fabrizio}},\ }\bibfield  {title} {{\selectlanguage {en}\bibinfo {title} {Luttinger's theorem in the presence of {Luttinger} surfaces}},\ }\href {https://doi.org/10.1103/PhysRevB.106.045109} {\bibfield  {journal} {\bibinfo  {journal} {Physical Review B}\ }\textbf {\bibinfo {volume} {106}},\ \bibinfo {pages} {045109} (\bibinfo {year} {2022})}\BibitemShut {NoStop}%
\bibitem [{\citenamefont {Yang}(2019)}]{yang_manifestation_2019}%
  \BibitemOpen
  \bibfield  {author} {\bibinfo {author} {\bibfnamefont {M.-F.}\ \bibnamefont {Yang}},\ }\bibfield  {title} {\bibinfo {title} {Manifestation of topological behaviors in interacting {Weyl} systems: {One}-body versus two-body correlations},\ }\href {https://doi.org/10.1103/PhysRevB.100.245137} {\bibfield  {journal} {\bibinfo  {journal} {Physical Review B}\ }\textbf {\bibinfo {volume} {100}},\ \bibinfo {pages} {245137} (\bibinfo {year} {2019})}\BibitemShut {NoStop}%
\bibitem [{\citenamefont {Meng}\ and\ \citenamefont {Budich}(2019)}]{meng_unpaired_2019}%
  \BibitemOpen
  \bibfield  {author} {\bibinfo {author} {\bibfnamefont {T.}~\bibnamefont {Meng}}\ and\ \bibinfo {author} {\bibfnamefont {J.~C.}\ \bibnamefont {Budich}},\ }\bibfield  {title} {\bibinfo {title} {Unpaired {Weyl} {Nodes} from {Long}-{Ranged} {Interactions}: {Fate} of {Quantum} {Anomalies}},\ }\href {https://doi.org/10.1103/PhysRevLett.122.046402} {\bibfield  {journal} {\bibinfo  {journal} {Physical Review Letters}\ }\textbf {\bibinfo {volume} {122}},\ \bibinfo {pages} {046402} (\bibinfo {year} {2019})}\BibitemShut {NoStop}%
\bibitem [{\citenamefont {Blesio}\ \emph {et~al.}(2018)\citenamefont {Blesio}, \citenamefont {Manuel}, \citenamefont {Roura-Bas},\ and\ \citenamefont {Aligia}}]{TQPTAndersonimpurity}%
  \BibitemOpen
  \bibfield  {author} {\bibinfo {author} {\bibfnamefont {G.~G.}\ \bibnamefont {Blesio}}, \bibinfo {author} {\bibfnamefont {L.~O.}\ \bibnamefont {Manuel}}, \bibinfo {author} {\bibfnamefont {P.}~\bibnamefont {Roura-Bas}},\ and\ \bibinfo {author} {\bibfnamefont {A.~A.}\ \bibnamefont {Aligia}},\ }\bibfield  {title} {\bibinfo {title} {Topological quantum phase transition between fermi liquid phases in an anderson impurity model},\ }\href {https://doi.org/10.1103/PhysRevB.98.195435} {\bibfield  {journal} {\bibinfo  {journal} {Phys. Rev. B}\ }\textbf {\bibinfo {volume} {98}},\ \bibinfo {pages} {195435} (\bibinfo {year} {2018})}\BibitemShut {NoStop}%
\bibitem [{\citenamefont {Žitko}\ \emph {et~al.}(2021)\citenamefont {Žitko}, \citenamefont {Blesio}, \citenamefont {Manuel},\ and\ \citenamefont {Aligia}}]{zitko_iron_2021}%
  \BibitemOpen
  \bibfield  {author} {\bibinfo {author} {\bibfnamefont {R.}~\bibnamefont {Žitko}}, \bibinfo {author} {\bibfnamefont {G.~G.}\ \bibnamefont {Blesio}}, \bibinfo {author} {\bibfnamefont {L.~O.}\ \bibnamefont {Manuel}},\ and\ \bibinfo {author} {\bibfnamefont {A.~A.}\ \bibnamefont {Aligia}},\ }\bibfield  {title} {{\selectlanguage {en}\bibinfo {title} {Iron phthalocyanine on {Au}(111) is a “non-{Landau}” {Fermi} liquid}},\ }\href {https://doi.org/10.1038/s41467-021-26339-z} {\bibfield  {journal} {\bibinfo  {journal} {Nature Communications}\ }\textbf {\bibinfo {volume} {12}},\ \bibinfo {pages} {6027} (\bibinfo {year} {2021})}\BibitemShut {NoStop}%
\bibitem [{\citenamefont {Kundu}\ and\ \citenamefont {Sénéchal}(2021)}]{kundu_spin_2021}%
  \BibitemOpen
  \bibfield  {author} {\bibinfo {author} {\bibfnamefont {S.}~\bibnamefont {Kundu}}\ and\ \bibinfo {author} {\bibfnamefont {D.}~\bibnamefont {Sénéchal}},\ }\bibfield  {title} {\bibinfo {title} {Spin density wave order in interacting type-{I} and type-{II} {Weyl} semimetals},\ }\href {https://doi.org/10.1103/PhysRevB.103.085136} {\bibfield  {journal} {\bibinfo  {journal} {Physical Review B}\ }\textbf {\bibinfo {volume} {103}},\ \bibinfo {pages} {085136} (\bibinfo {year} {2021})}\BibitemShut {NoStop}%
\end{thebibliography}%
\let\addcontentsline\oldaddcontentsline

\clearpage

\renewcommand{\t}[1]{\text{#1}}
\renewcommand{\theequation}{S\arabic{equation}}
\renewcommand{\selectlanguage}[1]{}
\renewcommand{\thefigure}{S\arabic{figure}}

\setcounter{equation}{0}

\renewcommand{\thesection}{S\arabic{section}}
\onecolumngrid
  \setcounter{table}{0}
\renewcommand{\thetable}{S\arabic{table}}%
\setcounter{figure}{0}

\begin{center}
  \textbf{\large Supplemental Material for ``The Weyl-Mott point:\ topological and non-Fermi liquid behavior from an isolated Green's function zero''}\\[.2cm]
R.\ Flores-Calderón$^{1,2,*}$ and C.\ Hooley$^{3}$\\[.1cm]
  {\itshape ${}^1$Max Planck Institute for the Physics of Complex Systems, Nöthnitzer Strasse 38, 01187 Dresden, Germany\\
  ${}^{2}$ Max Planck Institute for Chemical Physics of Solids, Nöthnitzer Strasse 40, 01187 Dresden, Germany\\
  ${}^{3}$Centre for Fluid and Complex Systems, Coventry University, Coventry CV1 2TT, United Kingdom} \\
  ${}^*$Electronic address:\ rflorescalderon@pks.mpg.de\\
(Dated \today)\\[1cm]
\end{center}

In this Supplemental Material we derive the form of the Green's function, the occupancy, and the spectral function as functions of the chemical potential, interaction strength, and hopping integral.

\tableofcontents

\section{Green's Function}

 We begin with the Hatsugai-Kohmoto-Weyl model in which the quadratic term is written in the spin-diagonalized basis but the interaction term is still in the original `up-down' basis:
\begin{align}
    H= \sum_{\textbf{k}} \lambda_{{\bf k}s}\alpha_{s\textbf{k}}^\dagger \alpha^{\phantom{\dagger}}_{s\textbf{k}}+U \sum_{\textbf{k}}n_{\textbf{k}\uparrow}n_{\textbf{k}\downarrow}.
\end{align}
The operators that we have used to diagonalize the quadratic part are 
\begin{align}
\alpha_{s\textbf{k}}=\sum_{\sigma}V_{s\sigma}(\textbf{k}) c_{\textbf{k}\sigma},
\end{align}
which may be inverted to give
\begin{align}
c_{\textbf{k}\sigma}=\sum_{s} \left[ V^\dagger(\textbf{k}) \right]_{\sigma s}\alpha_{\textbf{k}s}.
\end{align}
The matrix $V(\textbf{k})$  is formed from the eigenvectors of $h_0(\textbf{k})$, which means that $h_0(\textbf{k})=V^\dagger(\textbf{k}) D(\textbf{k}) V(\textbf{k})$ with $\left[ D(\textbf{k}) \right]_{ss'}=\delta_{ss'}\lambda_{s}(\textbf{k})$. Let us now change the basis for the Hatsugai-Kohmoto interaction term:
\begin{align}
    H_{\rm HK}=U \sum_{\textbf{k}}n_{\textbf{k}\uparrow}n_{\textbf{k}\downarrow}=U\sum_{\textbf{k}s,s'; l,l'} V^{\phantom{\dagger}}_{s\uparrow}V^\dagger_{\uparrow s'}V^{\phantom{\dagger}}_{l\downarrow}V^\dagger_{\downarrow l'}\alpha^\dagger_{\textbf{k}s}\alpha^{\phantom{\dagger}}_{\textbf{k}s'}\alpha^\dagger_{\textbf{k}l}\alpha^{\phantom{\dagger}}_{\textbf{k}l'},
\end{align}
where we have suppressed the explicit ${\bf k}$-dependence of $V$ for clarity.

This apparently complicated interaction term may be considerably simplified due to the unitarity of the diagonalizing operation.  Normal-ordering the operator using the fermionic anticommutation relation, $\alpha^{\phantom{\dagger}}_{\textbf{k}s'}\alpha^\dagger_{\textbf{k}l}=-\alpha^\dagger_{\textbf{k}l}\alpha^{\phantom{\dagger}}_{\textbf{k}s'}+\delta_{ls'}$, we obtain
\begin{align}
    H_{\rm HK}=U\sum_{\textbf{k}s,s';l} V^{\phantom{\dagger}}_{s\uparrow}V^\dagger_{\uparrow s'}V^{\phantom{\dagger}}_{s'\downarrow}V^\dagger_{\downarrow l}\alpha^\dagger_{\textbf{k}s}\alpha^{\phantom{\dagger}}_{\textbf{k}l}-U\sum_{\textbf{k}s,s'; l,l'} V^{\phantom{\dagger}}_{s\uparrow}V^\dagger_{\uparrow s'}V^{\phantom{\dagger}}_{l\downarrow}V^\dagger_{\downarrow l'}\alpha^\dagger_{\textbf{k}s}\alpha^\dagger_{\textbf{k}l}\alpha^{\phantom{\dagger}}_{\textbf{k}s'}\alpha^{\phantom{\dagger}}_{\textbf{k}l'}.
\end{align}
The first term is actually zero because of the unitarity property of $V$, which implies that $\sum_{s'}V^\dagger_{\sigma s'}V^{\phantom{\dagger}}_{\sigma'r'}=\delta_{\sigma,\sigma'}$.  The remaining term is
\begin{align}
    H_{\rm HK}=U\sum_{\textbf{k}s,s'; l,l'} V^{\phantom{\dagger}}_{s\uparrow}V^\dagger_{\uparrow s'}V^{\phantom{\dagger}}_{l\downarrow}V^\dagger_{\downarrow l'}\alpha^\dagger_{\textbf{k}l}\alpha^\dagger_{\textbf{k}s}\alpha^{\phantom{\dagger}}_{\textbf{k}s'}\alpha^{\phantom{\dagger}}_{\textbf{k}l'}=\sum_{\textbf{k}} U(\textbf{k}) m_{\textbf{k}+}m_{\textbf{k}-},
\end{align}
where $U(\textbf{k})/U=V^{\phantom{\dagger}}_{+\uparrow}V^\dagger_{\uparrow +}V^{\phantom{\dagger}}_{-\downarrow}V^\dagger_{\downarrow -}- V^{\phantom{\dagger}}_{-\uparrow}V^\dagger_{\uparrow +}V^{\phantom{\dagger}}_{+\downarrow}V^\dagger_{\downarrow -}+V^{\phantom{\dagger}}_{-\uparrow}V^\dagger_{\uparrow -}V^{\phantom{\dagger}}_{+\downarrow}V^\dagger_{\downarrow +}-V^{\phantom{\dagger}}_{+\uparrow}V^\dagger_{\uparrow -}V^{\phantom{\dagger}}_{-\downarrow}V^\dagger_{\downarrow +}$ and $m_{\textbf{k}\pm}=\alpha^\dagger_{\textbf{k}\pm}\alpha^{\phantom{\dagger}}_{\textbf{k}\pm}$.  $U(\textbf{k})$ may be simplified by using the unitarity of the matrix $V$:
\begin{align}
    U(\textbf{k})/U&=V^{\phantom{\dagger}}_{+\uparrow}V^\dagger_{\uparrow +}(1-V^{\phantom{\dagger}}_{-\uparrow}V^\dagger_{\uparrow -})-V^{\phantom{\dagger}}_{-\uparrow}V^\dagger_{\uparrow +}V^{\phantom{\dagger}}_{+\downarrow}V^\dagger_{\downarrow -}+V^{\phantom{\dagger}}_{-\uparrow}V^\dagger_{\uparrow -}(1-V^{\phantom{\dagger}}_{+\uparrow}V^\dagger_{\uparrow +})-V^{\phantom{\dagger}}_{+\downarrow}V^\dagger_{\downarrow -}V^{\phantom{\dagger}}_{-\uparrow}V^\dagger_{\uparrow +}\\
    &=V^{\phantom{\dagger}}_{+\uparrow}V^\dagger_{\uparrow +}+V^{\phantom{\dagger}}_{-\uparrow}V^\dagger_{\uparrow -}-2V^{\phantom{\dagger}}_{+\uparrow}V^\dagger_{\uparrow +}V^{\phantom{\dagger}}_{-\uparrow}V^\dagger_{\uparrow -}-2V^{\phantom{\dagger}}_{+\downarrow}V^\dagger_{\downarrow -}V^{\phantom{\dagger}}_{-\uparrow}V^\dagger_{\uparrow +}\\
    &=V^{\phantom{\dagger}}_{+\uparrow}V^\dagger_{\uparrow +}+V^{\phantom{\dagger}}_{-\uparrow}V^\dagger_{\uparrow -}-2V^{\phantom{\dagger}}_{+\uparrow}V^\dagger_{\uparrow -}V^\dagger_{\uparrow +}V^{\phantom{\dagger}}_{-\uparrow}-2V^{\phantom{\dagger}}_{+\downarrow}V^\dagger_{\downarrow -}V^{\phantom{\dagger}}_{-\uparrow}V^\dagger_{\uparrow +}\\
    &=V^{\phantom{\dagger}}_{+\uparrow}V^\dagger_{\uparrow +}+V^{\phantom{\dagger}}_{-\uparrow}V^\dagger_{\uparrow -}+2V^{\phantom{\dagger}}_{+\downarrow}V^\dagger_{\downarrow -}V^\dagger_{\uparrow +}V^{\phantom{\dagger}}_{-\uparrow}-2V_{+\downarrow}V^\dagger_{\downarrow -}V^{\phantom{\dagger}}_{-\uparrow}V^\dagger_{\uparrow +}\\
    &=V^{\phantom{\dagger}}_{+\uparrow}V^\dagger_{\uparrow +}+V^{\phantom{\dagger}}_{-\uparrow}V^\dagger_{\uparrow -}=\sum_{s}V^\dagger_{\uparrow s} V^{\phantom{\dagger}}_{s\uparrow}=1.
\end{align}
Hence our unitary transformation has given rise to the simpler Hamiltonian
\begin{align}
    H= \sum_{\textbf{k}} \lambda_{s}(\textbf{k})\alpha_{\textbf{k}s}^\dagger \alpha^{\phantom{\dagger}}_{\textbf{k}s}+ U \sum_{\textbf{k}} m_{\textbf{k}+}m_{\textbf{k}-}.
\end{align}

At each momentum $\textbf{k}$ there are only four states:\ $\ket{0},\ket{+},\ket{-},$ and $\ket{D}$, where $\pm$ indicates occupied $s=\pm$ states, $0$ is not occupied, and $D$ stands for double occupation.
We can write the energies for these four states as 
\begin{align}
    \ket{0}\rightarrow 0,\quad \ket{+}\rightarrow \lambda_{\textbf{k}-},\quad \ket{-}\rightarrow \lambda_{\textbf{k}+},\quad \ket{D}\rightarrow \lambda_{\textbf{k}+}+\lambda_{\textbf{k}-}+U(\textbf{k}).
\end{align}
We keep the $U(\textbf{k})$ notation although we proved it is just a constant. We can now proceed to calculate the simplest observable, the single-particle Green's function:
\begin{align}
    \tilde{G}_{\textbf{k}+}(\tau)&=-\expval{\hat {\mathcal{T}} \alpha^{\phantom{\dagger}}_{\textbf{k},+}(\tau)\alpha^\dagger_{\textbf{k},+}(0)}=-\expval{e^{\tau H}\alpha^{\phantom{\dagger}}_{\textbf{k},+}(0)e^{-\tau H}\alpha^\dagger_{\textbf{k},+}(0)}_T\\
    &=-\dfrac{1}{Z}\text{Tr}\{e^{(\tau-\beta) H}\alpha^{\phantom{\dagger}}_{\textbf{k},+}e^{-\tau H}\alpha^\dagger_{\textbf{k},+}\}=-\dfrac{1}{Z_\textbf{k}}(e^{-\tau \lambda_{\textbf{k}+}}+e^{-\tau(\lambda_{\textbf{k}+}+U(\textbf{k}))-\beta \lambda_{\textbf{k}-}}).
\end{align}
Let us write the average number of particles to compare with the previous studies and later on calculate the filling fraction
\begin{align}
    &\expval{m_{\textbf{k}s}}=\dfrac{1}{Z_{\textbf{k}}}(e^{-\beta \lambda_{\textbf{k}s}}+e^{-\beta (\lambda_{\textbf{k}+}+\lambda_{\textbf{k}-}+U(\textbf{k}))}),\\
    & 1-\expval{m_{\textbf{k}s}}=\dfrac{1}{Z_{\textbf{k}}}(1+e^{-\beta \lambda_{\textbf{k}\tilde{s}}}).
\end{align}
We can now go to Matsubara space and write the Green's function as
\begin{align}
    \tilde{G}_{\textbf{k}+}(i\omega_n)=\int_{0}^{\beta}\dd \tau \ \tilde{G}_{\textbf{k}+}(\tau)e^{i\omega_n \tau}.
\end{align}
We Fourier transform using the identity for the fermionic Matsubara frequencies $e^{i\omega_n \beta}=-1$, which implies that $\omega_n=(2n+1)\pi/\beta$:
\begin{align}
    \int_{0}^{\beta}\dd \tau \  e^{i\omega_n \tau} e^{-\lambda \tau}=\dfrac{1}{i\omega_n-\lambda}(-e^{-\beta \lambda }-1).
\end{align}
We can thus write the Matsubara Green's function as
\begin{align}
       \tilde{G}_{\textbf{k}+}(i\omega_n)&=\dfrac{1}{Z_{\textbf{k}}}\dfrac{1}{i\omega_n-\lambda_{\textbf{k}+}}(e^{-\beta \lambda_{\textbf{k}+} }+1)+\dfrac{1}{Z_{\textbf{k}}}\dfrac{e^{-\beta \lambda_{\textbf{k}-}}}{i\omega_n-(\lambda_{\textbf{k}+}+U(\textbf{k}))}(e^{-\beta (\lambda_{\textbf{k}+}+U(\textbf{k}))}+1)\\
       &=\dfrac{1-\expval{m_{\textbf{k}-}}}{i\omega_n-\lambda_{\textbf{k}+}}+\dfrac{\expval{m_{\textbf{k}-}}}{i\omega_n-(\lambda_{\textbf{k}+}+U(\textbf{k}))}.
\end{align}
We then find
\begin{align}
    \tilde{G}_{\textbf{k}s}(i\omega_n)=\dfrac{1-\expval{m_{\textbf{k}\bar{s}}}}{i\omega_n-\lambda_{\textbf{k}s}}+\dfrac{\expval{m_{\textbf{k}\bar{s}}}}{i\omega_n-(\lambda_{\textbf{k}s}+U(\textbf{k}))};
\end{align}
going back to the original fermions, we obtain
\begin{align}
     G_{\sigma\sigma'}(\textbf{k},i\omega_n)=\sum_{s}\left[ V(\textbf{k})^\dagger \right]_{\sigma s} \left(\dfrac{1-\expval{m_{\textbf{k}\bar{s}}}}{i\omega_n-\lambda_{\textbf{k}s}}+\dfrac{\expval{m_{\textbf{k}\bar{s}}}}{i\omega_n-(\lambda_{\textbf{k}s}+U(\textbf{k}))}\right)V_{s \sigma'}(\textbf{k}).
\end{align}

\begin{figure}[hb!]
    \centering
    \includegraphics[scale=0.5]{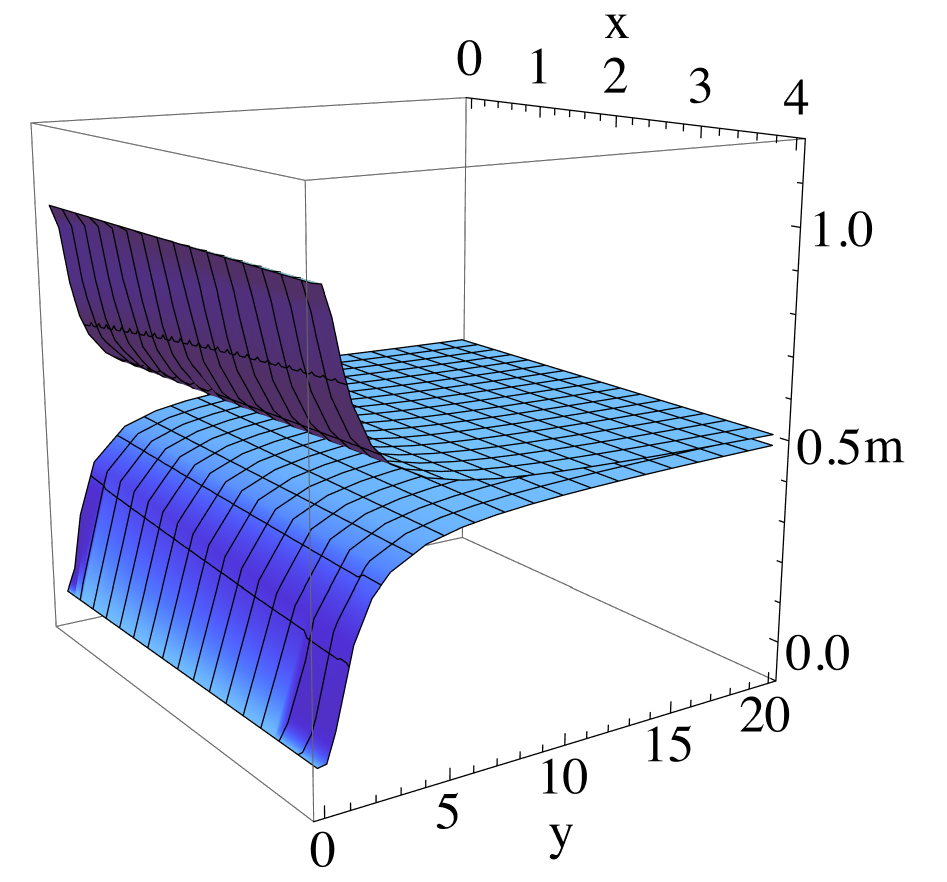}
    \caption{Half filling occupation $m_s(x,y)$ at each momentum except the Weyl points, where $y=\frac{T}{d_{\textbf{k}}},x=\frac{U}{d_{\textbf{k}}}$.}
    \label{fig:finitetemp}
\end{figure}

\section{Momentum occupation at half-filling: $\mu=U/2$}
Let us focus on the occupation now since we see it gives the weight of the Green's function. For a two-band Hamiltonian the energies are given by $\lambda_{\textbf{k}s}=-\mu+sd_{\textbf{k}}$, where $h_{\textbf{k}}=\vec{d}_{\textbf{k}}\cdot \vec{\sigma}-\mu \sigma_0$ and $s=\pm 1$. We notice that for $\mu=U/2$ we actually have
\begin{align}
    \expval{m_{\textbf{k}s}}=\dfrac{1}{Z_{\textbf{k}}}(e^{-\beta \lambda_{\textbf{k}s}}+e^{-\beta (-2\mu+U)})=\dfrac{1}{Z_{\textbf{k}}}(1+e^{-\beta \lambda_{\textbf{k}s}}).
\end{align}
The denominator can be easily calculated to be
\begin{align}
    Z_{\textbf{k}}= 2+\sum_{s'}e^{-\beta\lambda_{\textbf{k}s'}}.
\end{align}
We find then that the occupation for each momentum is fixed to half filling:
\begin{align}
    \expval{m_{\textbf{k}}}=\sum_s\expval{m_{\textbf{k}s}}=\dfrac{1}{ 2+\sum_{s'}e^{-\beta\lambda_{\textbf{k}s'}}}(\sum_s 1 +\sum_s e^{-\beta \lambda_{\textbf{k}s}})=1.
\end{align}
Moreover, the physical fermions also have the same occupation:
\begin{align}
  \expval{n_{\textbf{k}}} = \sum_{\sigma} \expval{n_{\textbf{k},\sigma}}=\sum_s \sum_{\sigma} V_{\sigma s}^\dagger \expval{m_{\textbf{k},s}}V_{s \sigma }=\sum_s \delta_{s,s}\expval{m_{\textbf{k},s}}=1.
\end{align}
This can be understood from the particle-hole symmetry of the model $\alpha^{\phantom{\dagger}}_{\textbf{k}s}\rightarrow b^{\phantom{\dagger}}_{\textbf{k}s}=\alpha^\dagger_{\textbf{k}\tilde{s}}$ at $\mu=U/2$. Now for this specific point we can write down
\begin{align}
    \expval{m_{\textbf{k}s}}=\dfrac{1+e^{-\beta \lambda_{\textbf{k}s}}}{2+e^{-\beta\lambda_{\textbf{k}-}}+e^{-\beta\lambda_{\textbf{k}+}}}.
\end{align}
Let us define $y=T/d_{\textbf{k}},x=U/d_{\textbf{k}}$, for $\textbf{k}\neq \pm \textbf{Q}$, where $\pm {\bf Q}$ are the Weyl points where $d_{\textbf{k}}=0$.  We can then rewrite the function in the argument of the exponential as $f_s(x,y)=\beta(-\mu+s d_{\textbf{k}})=\dfrac{s}{y}-\dfrac{x}{2y}$. The average occupation then looks like
\begin{align}
      \expval{m_{\textbf{k}s}}=m_s(x,y)=\dfrac{1+e^{-f_s(x,y)}}{2+e^{-f_-(x,y)}+e^{-f_+(x,y)}}.
\end{align}
Let us explore some of the limits of this function. If we focus on a finite interaction $U$ and take the limit of zero temperature then the signs of $f_s$ are important since for $s=-1$ we have that $-f_-=(x/2+1)/y$ will be always positive and $-f_+=(x/2-1)/y$ is positive if $x/2-1>0$ meaning $x>2$ or $U>2d_{\textbf{k}}\approx 2t$ otherwise for $x<2$ implying $-f_+<0$ and so $\expval{m_{\textbf{k}+}}=0$, by the previous sum rule $\expval{m_{\textbf{k}-}}=1$. Nevertheless for $x>2$ the exponentials have different exponents $a/y$ and $b/y$ with $a=x/2+1> b=x/2-1$ and so in the limit of $y\rightarrow 0$ the exponential with $a$ wins and again $\expval{m_{\textbf{k}-}}=1$. 

In Fig. \ref{fig:finitetemp} we plot $m_s(x,y)$ for a wider range of parameters meaning for finite temperature. We see that as soon as temperature $y\neq 0$ is turn on the occupations deviate from zero and one for all $d_{\textbf{k}}\neq 0$. This means that the bands which had only occupation zero get now occupied. The weak dependence observed as a funciton of $U$ or $x$ is related to the fact that the changes in occupation as a function of temperature behave exponentially.

\section{Green's function zeros and non-Fermi liquid behavior}
At the Weyl point we can write the Green's function exactly to see there is always for finite $U$ a divergent self-energy $\Sigma^R(\textbf{Q},\omega)$ for $\omega\rightarrow 0$, since
\begin{align}
    \tilde{G}_s(\textbf{Q},\omega)= \dfrac{1}{\omega+i0^+-\dfrac{(U/2)^2}{\omega+i0^+}}.
\end{align}
We conclude that for any finite $U$ at half-filling the resulting state cannot be adiabatically connected to the original non-interacting Weyl semimetal. Not only that but we see the simplest case of a Luttinger-Lifshitz transition since a single pole of the Green's function at the Weyl point for $U=0$ gets transformed into a single zero of the Green's function also at the Weyl point
\begin{align}
     \tilde{G}_s(\textbf{Q},\omega\rightarrow 0)= \lim_{\omega\rightarrow 0}\left\{\dfrac{1}{\omega+i0^+ +U/2}+\dfrac{1}{\omega+i0^+-U+U/2}\right\}=0.
\end{align}
For calculating the zeros of the physical Green's function we would need to solve $G=0$ meaning
\begin{align}
     &G_{\sigma\sigma'}(\textbf{k},\omega)=\sum_{s}V_{\sigma s}(\textbf{k})^\dagger\left(\dfrac{1-\expval{m_{\textbf{k}\bar{s}}}}{\omega+i0^+-\lambda_{\textbf{k}s}}+\dfrac{\expval{m_{\textbf{k}\bar{s}}}}{\omega+i0^+-(\lambda_{\textbf{k}s}+U(\textbf{k}))}\right)V_{s \sigma'}(\textbf{k})=0\\
      &G_{\sigma\sigma'}(\textbf{k},\omega)=V_{\sigma +}(\textbf{k})^\dagger V_{+ \sigma'}(\textbf{k}) \dfrac{1}{\omega+i0^+-(\lambda_{\textbf{k}+}+U)}+V_{\sigma -}(\textbf{k})^\dagger V_{- \sigma'}(\textbf{k})\dfrac{1}{\omega+i0^+-\lambda_{\textbf{k}-}}=0\\
      &V_{\sigma +}(\textbf{k})^\dagger V_{+ \sigma'}(\textbf{k})(\omega+i0^+-\lambda_{\textbf{k}-})=-V_{\sigma -}(\textbf{k})^\dagger V_{- \sigma'}(\textbf{k})(\omega+i0^+-\lambda_{\textbf{k}+}-U)\\
      &\delta_{\sigma\sigma'}\omega = V_{\sigma +}(\textbf{k})^\dagger V_{+ \sigma'}(\textbf{k}) \lambda_{\textbf{k}-} + V_{\sigma -}(\textbf{k})^\dagger V_{- \sigma'}(\textbf{k})(\lambda_{\textbf{k}+}+U),
\end{align}
where we used that $V^\dagger V = \mathbb{I}$. We conclude there are two branches with zeros of the Green's function given by
\begin{align}
    &\omega_{\uparrow}=V_{\uparrow +}(\textbf{k})^\dagger V_{+ \uparrow}(\textbf{k}) \lambda_{\textbf{k}-} + V_{\uparrow -}(\textbf{k})^\dagger V_{- \uparrow}(\textbf{k})(\lambda_{\textbf{k}+}+U),\\
    & \omega_{\downarrow}=V_{\downarrow +}(\textbf{k})^\dagger V_{+ \downarrow}(\textbf{k}) \lambda_{\textbf{k}-} + V_{\downarrow -}(\textbf{k})^\dagger V_{- \downarrow}(\textbf{k})(\lambda_{\textbf{k}+}+U).
\end{align}
For the 3D model we simply plot the determinant of the Green's function in Fig. 2 of the main text, since it will indicate poles when it diverges and zeros when it is zero. Remarkably in this quantity we see the Weyl points are indeed still connected. At zero frequency we have a single zero of the Green's function.

Let us consider a generic zero of the retarded diagonal Green's function and place it at $\omega=\mu$ and look at the Weyl momenta \textbf{Q}
\begin{align}
    &\dfrac{1-\expval{m_{\textbf{k}\bar{s}}}}{\omega+i0^+-\lambda_{\textbf{k}s}}+\dfrac{\expval{m_{\textbf{k}\bar{s}}}}{\omega+i0^+-(\lambda_{\textbf{k}s}+U)}=0.\\
    &\dfrac{1-\expval{m_{\textbf{k}\bar{s}}}}{\omega+i0^+-\lambda_{\textbf{k}s}}=-\dfrac{\expval{m_{\textbf{k}\bar{s}}}}{\omega+i0^+-(\lambda_{\textbf{k}s}+U)}\\
    &\dfrac{1}{\expval{m_{\textbf{Q}\bar{s}}}-1}\left(\mu+i0^+-(\lambda_{\textbf{Q}s})\right)=\dfrac{1}{\expval{m_{\textbf{Q}\bar{s}}}}\left(\mu+i0^+-(\lambda_{\textbf{Q}s}+U)\right)\\
    &\dfrac{1}{\expval{m_{\textbf{Q}\bar{s}}}-1}\left(\mu+i0^+-(-\mu)\right)=\dfrac{1}{\expval{m_{\textbf{Q}\bar{s}}}}\left(\mu+i0^+-(-\mu+U)\right)\\
    &\dfrac{1}{\expval{m_{\textbf{Q}\bar{s}}}-1}2\mu=\dfrac{1}{\expval{m_{\textbf{Q}\bar{s}}}}\left(2\mu-U)\right)\\
    &\expval{m_{\textbf{Q}\bar{s}}}=(\expval{m_{\textbf{Q}\bar{s}}}-1)\left(1-\dfrac{U}{2\mu}\right)\\
    &\expval{m_{\textbf{Q}\bar{s}}}=\expval{m_{\textbf{Q}\bar{s}}}\left(1-\dfrac{U}{2\mu}\right)-\left(1-\dfrac{U}{2\mu}\right)\\
    & 0=-\expval{m_{\textbf{Q}\bar{s}}}\dfrac{U}{2\mu}-\left(1-\dfrac{U}{2\mu}\right)
    \\
    & \expval{m_{\textbf{Q}\bar{s}}}=\left(1-\dfrac{2\mu}{U}\right).
\end{align}
Because we are talking about fermions we require
$0\leq\left(1-\dfrac{2\mu}{U}\right)\leq1 $ so $\mu>0,U>0$ and $2\mu\leq U$. The occupation at the Weyl point is
\begin{align}
    &\expval{m_{\textbf{Q}\bar{s}}}=\dfrac{1}{Z_{\textbf{Q}}}(e^{-\beta \lambda_{\textbf{Q}s}}+e^{-\beta (\lambda_{\textbf{Q}+}+\lambda_{\textbf{Q}-}+U)})\\
    &= \dfrac{1}{1+e^{-\beta \lambda_{\textbf{Q}-}}+e^{-\beta \lambda_{\textbf{Q}+}}+e^{-\beta (\lambda_{\textbf{Q}+}+\lambda_{\textbf{Q}-}+U)}}(e^{\beta \mu}+e^{-\beta (-2\mu+U)})
    = \dfrac{e^{\beta \mu}+e^{-\beta (-2\mu+U)}}{1+e^{\beta \mu}+e^{\beta \mu}+e^{-\beta (-2\mu+U)}}\\
    &= \dfrac{1+e^{\beta (\mu-U)}}{e^{-\beta \mu}+2+e^{\beta (\mu-U)}},
\end{align}
so that at zero temperature
\begin{align}
    \left(1-\dfrac{2\mu}{U}\right)=\dfrac{1}{2},\\
    \dfrac{U}{4}=\mu.
\end{align}
We can also calculate the occupation of the Weyl-Mott point at critical chemical potentials such as $\mu=0$ for which
\begin{align}
    & \lim_{\beta \rightarrow \infty} \lim_{\mu\rightarrow 0} \expval{m_{\textbf{Q}\bar{s}}}=  \lim_{\beta \rightarrow \infty} \lim_{\mu\rightarrow 0} \dfrac{e^{\beta \mu}+e^{-\beta (-2\mu+U)}}{1+e^{\beta \mu }+e^{\beta \mu}+e^{-\beta (-2\mu+U)}}=  \lim_{\beta \rightarrow \infty} \dfrac{1+e^{-\beta U}}{3+e^{-\beta U}}=\dfrac{1}{3};
\end{align}
similarly we have for $\mu=U$
\begin{align}
    & \lim_{\beta \rightarrow \infty} \lim_{\mu\rightarrow U} \expval{m_{\textbf{Q}\bar{s}}}=  \lim_{\beta \rightarrow \infty} \lim_{\mu\rightarrow 0} \dfrac{e^{\beta \mu}+e^{-\beta (-2\mu+U)}}{1+e^{\beta \mu }+e^{\beta \mu}+e^{-\beta (-2\mu+U)}}=  \lim_{\beta \rightarrow \infty} \dfrac{e^{\beta U}+e^{\beta U}}{1+e^{\beta U}+e^{\beta U}+e^{\beta U}}=\dfrac{2}{3}.
\end{align}
At the point $\mu=0$ the Green's function becomes 

\begin{align}
    \tilde{G}_{\textbf{k}_0s}(\omega)=\dfrac{1-1/3}{\omega+i0^+-(-\mu)}+\dfrac{1/3}{\omega+i0^+ -(-\mu+U)}, \quad \mu=0\\
    \tilde{G}_{\textbf{k}_0s}(\omega)=\dfrac{2/3}{\omega+i0^+}+\dfrac{1/3}{\omega+i0^+ -U}, \quad \mu=0
\end{align}
Equating the right hand side to zero we find then that the Green's function zero is now at 
\begin{align}
    \omega=\dfrac{2U}{3}, \quad \mu=0
\end{align}

At the point $\mu=U$ the Green's function becomes 

\begin{align}
    \tilde{G}_{\textbf{k}_0s}(\omega)=\dfrac{1-2/3}{\omega+i0^+-(-\mu)}+\dfrac{2/3}{\omega+i0^+ -(-\mu+U)}, \quad \mu=U\\
    \tilde{G}_{\textbf{k}_0s}(\omega)=\dfrac{1/3}{\omega+i0^+ +U}+\dfrac{2/3}{\omega+i0^+ }, \quad \mu=U
\end{align}
Equating the right hand side to zero we find then that the Green's function zero is now at 
\begin{align}
    \omega=-\dfrac{2U}{3}, \quad \mu=U
\end{align}

\section{General filling at zero temperature}\label{SGeneralfilling}
We calculate the properties of the model away from $\mu=U/2$ but for $U>0$:
\begin{align}
    \expval{m_{\textbf{k}s}}=\dfrac{1}{Z_{\textbf{k}}}(e^{-\beta \lambda_{\textbf{k}s}}+e^{-\beta (-2\mu+U)})=\dfrac{e^{-\beta \lambda_{\textbf{k}s}}+e^{-\beta (-2\mu+U)}}{1+e^{-\beta \lambda_{\textbf{k}\bar{s}}}+e^{-\beta \lambda_{\textbf{k}s}}+e^{-\beta (-2\mu+U)}}.
\end{align}
The total occupation will be
\begin{align}
\expval{n}=&\sum_{\textbf{k},\sigma}\expval{n_{\textbf{k}\sigma}}=\sum_{\textbf{k},s}\expval{m_{\textbf{k}s}}=\sum_{\textbf{k},s}\dfrac{e^{-\beta \lambda_{\textbf{k}s}}+e^{-\beta (-2\mu+U)}}{1+e^{-\beta \lambda_{\textbf{k}\bar{s}}}+e^{-\beta \lambda_{\textbf{k}s}}+e^{-\beta (-2\mu+U)}}\\
    =&\sum_{\textbf{k}}\dfrac{e^{-\beta \lambda_{\textbf{k}-}}+e^{-\beta \lambda_{\textbf{k}+}}+2e^{-\beta (-2\mu+U)}}{1+e^{-\beta \lambda_{\textbf{k}+}}+e^{-\beta \lambda_{\textbf{k}-}}+e^{-\beta (-2\mu+U)}}=\sum_{\textbf{k}}\dfrac{e^{-\beta (-\mu-d_{\textbf{k}})}+e^{-\beta (-\mu+d_{\textbf{k}})}+2e^{-\beta (-2\mu+U)}}{1+e^{-\beta \lambda_{\textbf{k}+}}+e^{-\beta \lambda_{\textbf{k}-}}+e^{-\beta (-2\mu+U)}}.
\end{align}
We note that $d_{\textbf{k}}>0$ always except at the Weyl points.  To proceed further we need to classify several cases.

\subsection{$\mu>0$}
We consider positive $\mu>0$ then $-\lambda_{\textbf{k}-}>-\lambda_{\textbf{k}+}$ equivalently if $\mu+d_{\textbf{k}}>-(-\mu+d_{\textbf{k}})=\mu-d_{\textbf{k}}$  which is always true since $d_{\textbf{k}}>0$ away from the Weyl point. These means that the denominator of 
\begin{align}
    \expval{m_{\textbf{k}+}}=\dfrac{e^{-\beta (-\mu+d_{\textbf{k}})}+e^{-\beta (-2\mu+U)}}{1+e^{-\beta \lambda_{\textbf{k}+}}+e^{-\beta \lambda_{\textbf{k}-}}+e^{-\beta (-2\mu+U)}},
\end{align}
will have a more diverging exponent than the numerator always unless $-(-2\mu+U)>-\lambda_{\textbf{k}-}$ or equivalently $2\mu-U>\mu+d_{\textbf{k}}$  which will be true if $\mu>U+d_{\textbf{k}}$. This means that the region in momentum space with $\mu-U<d_{\textbf{k}}$ and $\mu>U$ will have $\expval{m_{\textbf{k}+}}=0$. While the rest of momentum space has the same divergence in numerator and denominator giving $\expval{m_{\textbf{k}+}}=1$. We call these two regions $\mathcal{R}^+_<=\{\textbf{k}| \ d_{\textbf{k}}<\mu-U\}$ and $\mathcal{R}^+_>=\{\textbf{k}| \ d_{\textbf{k}}>\mu-U\}$. We can then collect the results for $\expval{m_{\textbf{k}+}}$ and $\mu>U$.
\begin{align}
    \expval{m_{\textbf{k}+}}=1, \quad \textbf{k}\in \mathcal{R}^+_<\\
    \expval{m_{\textbf{k}+}}=0, \quad \textbf{k}\in \mathcal{R}^+_>
\end{align}
Given that $d_{\textbf{k}}$ has an upper bound $W\propto 2t$ we know that near a Weyl point where $d_{\textbf{k}}=0$ we will nucleate electron pockets in the $\mathcal{R}^+_<$ region if $\mu>U$. This regions will have an volume of $4\pi(\mu-U)^3/3$. 
For the other component we note that if $-(-2\mu+U)$ dominates over all other exponents then $\expval{m_{\textbf{k}-}}=1$ since numerator and denominator cancel in the limit of zero temperature. Otherwise we have $-\lambda_{\textbf{k}-}$ dominates and again we have both numerator and denominator cancelling although it is a different exponent which will give  $\expval{m_{\textbf{k}-}}=1$ ,$\expval{m_{\textbf{k}+}}=0$.  \\

So we conclude that there are only two values of the occupancy in the ground state at $T=0$ given by half-filling if $0<\mu<U$ and fully-filled if $\mu>U>0$.
\begin{align}
     &\dfrac{\expval{n}}{V} =1, \quad 0<\mu<U,\\
     &\dfrac{\expval{n}}{V} =2, \quad U<\mu.
\end{align}

\subsection{$\mu<0$}
\begin{figure}[ht!]
    \centering
    \includegraphics[scale=0.5]{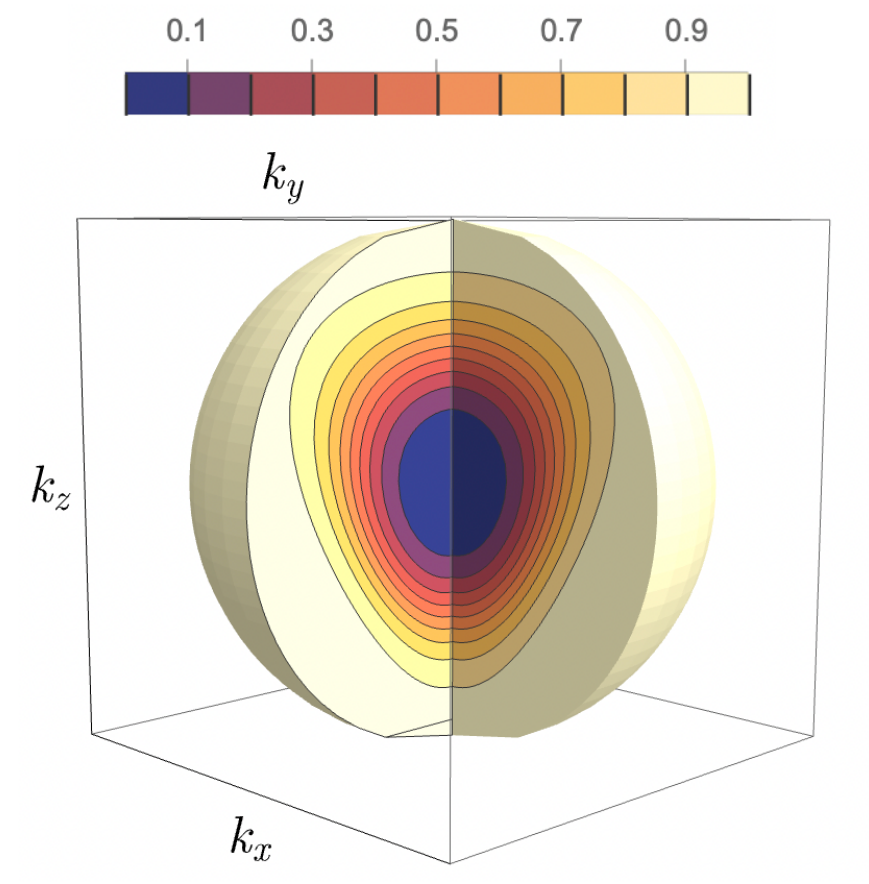}
    \caption{Occupation of the negative branch close to a Weyl point for $M=2, \beta = 10, \mu=-0.5, U= 2$}
    \label{fighole}
\end{figure}
This case is more subtle than the last since now $-\mu+d_{\textbf{k}}>0$ and $-\mu-d_{\textbf{k}}$ can be positive or negative depending on the momentum. Let us call $\mathcal{R}_<$ the region of momentum for which $-\mu-d_{\textbf{k}}>0$ and $\mathcal{R}_>$ the region for $-\mu-d_{\textbf{k}}<0$. If we look at the occupation of the negative branch we would obtain
\begin{align}
    \expval{m_{\textbf{k}-}}=\dfrac{e^{-\beta (-\mu-d_{\textbf{k}})}+e^{-\beta (-2\mu+U)}}{1+e^{-\beta \lambda_{\textbf{k}+}}+e^{-\beta \lambda_{\textbf{k}-}}+e^{-\beta (-2\mu+U)}},
\end{align}
For $\mu<0$ we have $-(-2\mu+U)=2\mu-U<0$ always which means the second factor in the numerator goes to zero for $\beta\rightarrow \infty$. If $-(-\mu-d_{\textbf{k}})<0$  which is region $\mathcal{R}_<$ then the whole numerator goes to zero and since $-(-\mu+d_{\textbf{k}})<0$ always then we get a single one in the denominator and a zero in the numerator. For the opposite region we have $-(-\mu-d_{\textbf{k}})>0$ which then diverges for $\beta\rightarrow \infty$ and both numerator and denominator scale to the same number giving one. All together we have
\begin{align}
    \expval{m_{\textbf{k}-}}=0, \quad \textbf{k}\in \mathcal{R}_<\\
    \expval{m_{\textbf{k}-}}=1, \quad \textbf{k}\in \mathcal{R}_>
\end{align}
where we defined $\mathcal{R}_<=\{\textbf{k}| \ d_{\textbf{k}}<\abs{\mu}\}$ and $\mathcal{R}_>=\{\textbf{k}| \  d_{\textbf{k}}>\abs{\mu}\}$. We can interpret then $\mathcal{R}_<$ as hole pockets or regions of momentum space where the occupation is zero. Clearly near a Weyl point where $d_{\textbf{k}}=0$ we would have momentum in $\mathcal{R}_<$ so this points nucleates a hole pocket. An example for finite temperature is shown in Fig.~\ref{fighole}. Now since $d_{\textbf{k}}$ has an upper bound, given by the energy width $W\propto 2t$, if we lower the chemical potential enough at the critical value of $\abs{\mu}=W$ there will be no momentum in the BZ belonging to $R_>$. This means that the nucleation is complete and the system is empty.

\section{Spectral function}

We calculate now the spectral function for such $\expval{m_{\textbf{k}-}}=1$, $\expval{m_{\textbf{k}+}}=0$ momentum, which we show occurs generically for $0<\mu<U$:
\begin{align}
    \tilde{A}_s(\textbf{k},\omega) = -2 \Im\{\tilde{G}_{\textbf{k}s}(i\omega_n \rightarrow \omega+i0^+)\}=2\pi \left( (1-\expval{m_{\textbf{k}\bar{s}}})\delta(\omega-\lambda_{\textbf{k}s})+\expval{m_{\textbf{k}\bar{s}}}\delta(\omega-\lambda_{\textbf{k}s}-U)\right).
\end{align}
Plugging in the occupation values we get
\begin{align}
    &\tilde{A}_+(\textbf{k},\omega)=2\pi \delta(\omega-\lambda_{\textbf{k}+}-U), \qquad \qquad \tilde{A}_-(\textbf{k},\omega)=2\pi \delta(\omega-\lambda_{\textbf{k}-}),\\
    &\tilde{A}_+(\textbf{k},\omega)=2\pi \delta(\omega-(d_{\textbf{k}}+U-\mu)), \qquad  \tilde{A}_-(\textbf{k},\omega)=2\pi \delta(\omega+(d_{\textbf{k}}+\mu)),
\end{align}
while at the Weyl points we always have $\expval{m_{\textbf{k}\bar{s}}}=1/2$ so that
\begin{align}
    \tilde{A}_s(\textbf{Q},\omega) =\pi \left( \delta(\omega+U/2)+\delta(\omega-U/2)\right).
\end{align}
Of course, the physical spectral function which has a spin index mixes both, for example
\begin{align}
    &A_{\uparrow\uparrow}(\textbf{k},\omega)=-2 \Im\{G_{\textbf{k}\uparrow\uparrow}(i\omega_n \rightarrow \omega+i0^+)\}\\
    &= 2 \pi \left( V_{\uparrow+}^\dagger(\textbf{k})V_{+\uparrow}(\textbf{k})\delta(\omega-d_{\textbf{k}}-U/2)+(1-V_{\uparrow+}^\dagger(\textbf{k})V_{+\uparrow}(\textbf{k}))\delta(\omega+d_{\textbf{k}}+U/2)\right).
\end{align}
Now for the special points $\mu=0$ and $\mu=U$ we have still the same behaviour away from the Weyl point but at the Weyl point we have $\expval{m_{\textbf{k}\bar{s}}}=1/3$ and $\expval{m_{\textbf{k}\bar{s}}}=2/3$ respectively. Meaning then a spectral function of the form
\begin{align}
    \tilde{A}_s(\textbf{k},\omega) =2\pi \left( (1-1/3)\delta(\omega+\mu-s d_{\textbf{k}_0})+(1/3)\delta(\omega+\mu-s d_{\textbf{k}_0}-U)\right), \quad \mu=0\\
    \tilde{A}_s(\textbf{k}_0,\omega) =2\pi \left( (1-2/3)\delta(\omega+\mu-s d_{\textbf{k}_0})+(2/3)\delta(\omega+\mu-s d_{\textbf{k}_0}-U)\right), \quad \mu=U
\end{align}
Simplifying we get

\begin{align}
    \tilde{A}_s(\textbf{k}_0,\omega) =\dfrac{2\pi}{3} \left( 2\delta(\omega)+\delta(\omega-U)\right), \quad \mu=0\\
    \tilde{A}_s(\textbf{k}_0,\omega) =\dfrac{2\pi}{3} \left( \delta(\omega+U)+2\delta(\omega)\right), \quad \mu=U
\end{align}
We thus find spectral weight at $\omega=0$ characteristic of a metallic phase but with a two pole structure at the Weyl-Mott point, indeed this points are then non-Fermi liquids.
\section{Luttinger count}

We can further characterize how far our system is from the non-interacting case by calculating the Luttinger count, defined as \cite{sachdev_chern_2023}
\begin{align}
    N_1[G]=-\dfrac{1}{2\pi i} \int \dd z \ e^{z0^+}\Tr[G^{-1}(z)\dfrac{\partial G(z)}{\partial z}].
\end{align}
Because of the trace properties and the independence on frequency of $V(\textbf{k})$ we can calculate this from the diagonal Green's function
\begin{align}
    N_1[G]&=-\dfrac{1}{2\pi i} \int \dd z \ e^{z0^+}\sum_s\tr[\tilde{G}_{s}^{-1}(z)\dfrac{\partial \tilde{G}_{s}(z)}{\partial z}].
\end{align}
We find then that the total Luttinger count reduces to the sum of the Luttinger count for each separate diagonal species. Therefore we can calculate it with the usual formula valid at zero temperature, which is
\begin{align}
     n_1=\dfrac{1}{L^3}\sum_{\textbf{k}s}\Theta\left[\Re(\tilde{G}_s(\textbf{k},0))\right].
\end{align}
Then we need
\begin{align}
    &\Re(\tilde{G}_s(\textbf{k},0))=\Re\left(\dfrac{1-\expval{m_{\textbf{k}\bar{s}}}}{i0^+-\lambda_{\textbf{k}s}}+\dfrac{\expval{m_{\textbf{k}\bar{s}}}}{i0^+-(\lambda_{\textbf{k}s}+U)}\right)>0;\\
    &\Re\dfrac{1-\expval{m_{\textbf{k}\bar{s}}}}{i0^+-\lambda_{\textbf{k}s}}>-\Re \dfrac{\expval{m_{\textbf{k}\bar{s}}}}{i0^+-(\lambda_{\textbf{k}s}+U)};\\
     &\Re\dfrac{1-\expval{m_{\textbf{k}\bar{s}}}}{i0^+-\lambda_{\textbf{k}s}}>\Re \dfrac{\expval{m_{\textbf{k}\bar{s}}}}{i0^-+(\lambda_{\textbf{k}s}+U)}.\\
\end{align}
From the previous calculations we note that this inequality implies for $s=+$ that $d_{\textbf{k}}<\mu-U<0$ if $0<\mu<U$ which is impossible while for $s=-$ it is satisfied if $-\mu-d_\textbf{k}<0$ so it is always satisfied; we thus obtain
\begin{align}
    n_1= \dfrac{1}{L^3}\sum_{\textbf{k}} = 1, \qquad 0<\mu<U.
\end{align}

\end{document}